\documentclass[runningheads]{llncs}
\usepackage{graphicx}
\usepackage{subfigure}
\usepackage{amsmath}

\usepackage{booktabs}

\usepackage{appendix}

\usepackage[linesnumbered,ruled,vlined]{algorithm2e} 

\begin{document}
\title{Causality-Aware Neighborhood Methods for Recommender Systems}

\author{Masahiro Sato \and
	Janmajay Singh \and
	Sho Takemori \and
	Qian Zhang}


\authorrunning{M. Sato et al.}

\institute{Fuji Xerox, Yokohama, Japan 
	\email{\{sato.masahiro,janmajay.singh,takemori.sho,qian.zhang\}@fujixerox.co.jp}} 

\maketitle              

\begin{abstract}
	The business objectives of recommenders, such as increasing sales, are aligned with the causal effect of recommendations.
	Previous recommenders targeting for the causal effect employ the inverse propensity scoring (IPS) in causal inference.
	However, IPS is prone to suffer from high variance.
	The matching estimator is another representative method in causal inference field.
	It does not use propensity and hence free from the above variance problem.
	In this work, we unify traditional neighborhood recommendation methods with the matching estimator, and develop robust ranking methods for the causal effect of recommendations.
	Our experiments demonstrate that the proposed methods outperform various baselines in ranking metrics for the causal effect.
	The results suggest that the proposed methods can achieve more sales and user engagement than previous recommenders.
\end{abstract}

\section{Introduction}
\begin{figure}[bp]
	\centering
	\includegraphics[width=0.8\textwidth]{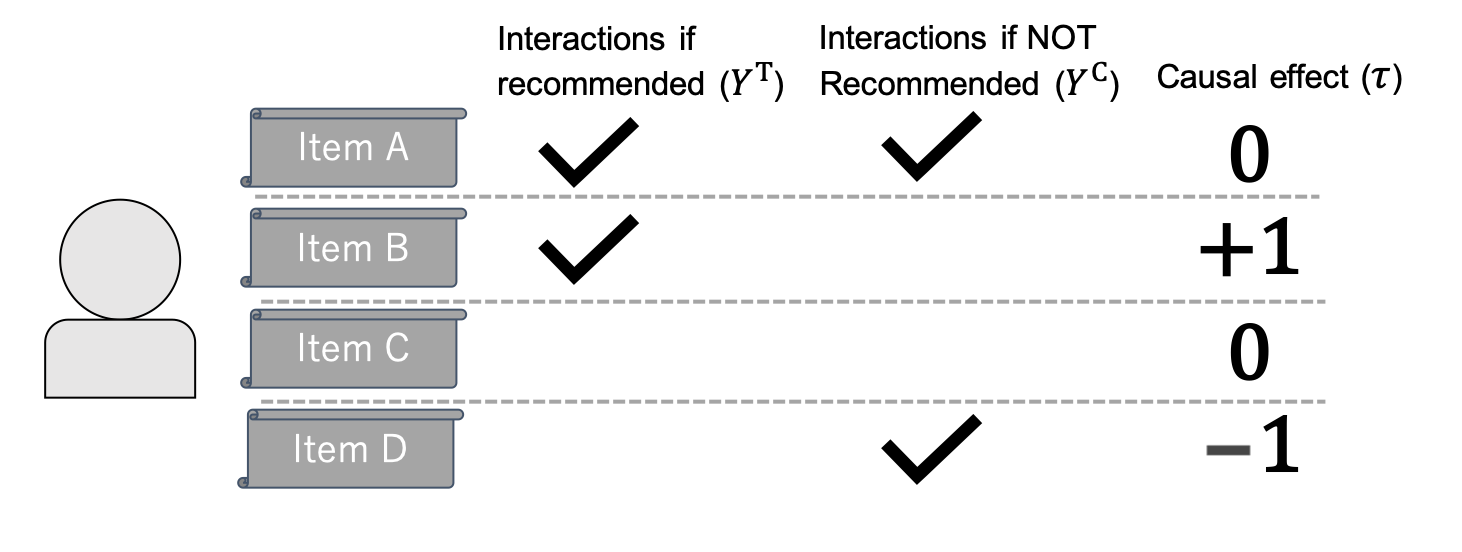}
	\caption{A figure to illustrate the causal effect of recommendations. 
		Recommending Item B results in increase of user interactions than without recommending, hence it has positive causal effect.} 
	\label{fig:concept}
\end{figure}

Recommender systems have been used in various services to improve sales and user engagement \cite{Jannach19}.
For these purposes, it is essential to increase users' positive interactions, such as purchases and views.
If recommended items are purchased or viewed, the recommendations are typically considered to be successful.
However, the recommended items might have been interacted even without the recommendations.
In this case, the user interactions are not \textit{caused} by the recommendations.
For example, if a user is an enthusiastic fan of a movie director, the user would watch a new movie of the director whether it is recommended or not.
Sharma et al.~\cite{Sharma15} analyzed the browsing logs of an e-commerce site and revealed that at least 75\% of recommended visits would likely occur in the absence of the recommendations.
To improve sales and user engagement, it is important to generate recommendations that truly increase user interactions.

Such an increase produced purely by recommendation is called \textit{causal effect}.
Fig. \ref{fig:concept} illustrates the causal effect of recommendation.
It is the difference of user interactions in two cases: if recommended and if not recommended.
The challenge for ranking items by the causal effect is that we can not directly measure the causal effect since an item is either recommended or not for a specific user.
Such unobservable nature is a fundamental problem of causal inference~\cite{Holland86} and various methods have been developed to address the problem~\cite{Imbens15,Hernan20}.

Few works targeting recommendation causal effect exist~\cite{Bodapati08,Sato16,Sato19,Sato20}, and it is largely an unexplored area of research.
Among them, a recent work~\cite{Sato20} employed IPS method~\cite{Lunceford04} in causal inference field, and developed unbiased learning-to-rank methods.
However, the IPS has been known to suffer from high variance due to small propensities~\cite{Swaminathan15b,Wang19,Saito20b}.
Although the previous work~\cite{Sato20} mitigates the variance by propensity capping, it incurs bias and affects the recommendation performance.
The matching estimator~\cite{Stuart10} is another representative method in causal inference.
It does not rely on propensities and enables a stable estimate of causal effect under various conditions of propensities.
Despite the potential advantage, there have been no attempts to apply the matching estimator for the causal effect of recommendations.

In this work, we explore the matching estimator approach to rank items by the causal effect of recommendations.
Matching estimators estimate causal effect by comparing observed outcomes for treated/untreated persons to those of similar persons in untreated/treated group.
Leveraging person similarity is analogous to traditional neighborhood recommendation methods.
We unify neighborhood recommendation methods with the matching estimator, and construct estimators of the causal effect for each user-item pair.
To obtain item rankings robust to randomness of user behaviors, we further improve the estimators by 1) mixing own and neighbor observations and 2) introducing a shrinkage hyper-parameter to adjust outcome estimates depending on computed neighborhood size.
We experimentally compare our methods with various baselines including recent IPS-based methods.
The results demonstrate the effectiveness of our methods for ranking items by the causal effect.
Such ranking can lead to increase of sales and user engagement, and have a practical benefit for businesses.

\section{Related Work}
Collaborative filtering is a widely used technique in recommender systems.
It can be grouped into the two general classes: neighborhood and model-based methods~\cite{Ning15,Koren15}.
Among model-based methods, matrix factorization models have been most popular~\cite{Koren09,Johnson14} and recently neural network models are gaining popularity~\cite{Zhang19}.
Neighborhood methods have been used since the dawn of recommender systems~\cite{Shardanand95,Sarwar01}.
They are still competitive to recent neural model-based methods~\cite{Dacrema19}, especially in session-based recommendation~\cite{Ludewig19}.
In this work, we extend neighborhood methods for the causal effect of recommendations.

Early work of recommendation for the causal effect proposed a two-stage purchase prediction model comprising awareness and satisfaction~\cite{Bodapati08}, similar to recent exposure modeling~\cite{Liang16}.
It assumed that recommendations make users aware of the items.
Later work~\cite{Sato16} incorporated user- and item-dependent responsiveness to recommendations.
Both methods predict purchase probabilities with and without recommendations and rank items by the difference of these probabilities.
Another strategy is to directly optimize ranking models for the causal effect~\cite{Sato19,Sato20}.
ULRMF and ULBPR~\cite{Sato19} are heuristic pointwise and pairwise learning methods inspired by the label transformation~\cite{Jaskowski12,Kane14} in uplift modeling~\cite{Radcliffe11,Devriendt18}.
Very recent work~\cite{Sato20} proposed DLCE, an IPS-based unbiased learning-to-rank method for the causal effect.

The IPS has been gaining popularity in counterfactual learning~\cite{Joachims16,Oosterhuis20}.
It has been applied to address missing not at random recommender feedback~\cite{Schnabel16,Saito20a}, position bias in information retrieval~\cite{Wang16,Joachims17,Agarwal19}, and selection bias in bandit feedback~\cite{Bottou13,Zhuang20}.
Domain adaptation is another counterfactual learning method~\cite{Johansson16} that are applied for recommenders~\cite{Bonner18}.
To the best of our knowledge, matching estimator has not been applied for recommenders or information retrieval.

\section{Preliminaries}
\subsection{Matching Estimator for Causal Inference}
\label{subsec:matching_estimator}
Let $Y_{n}^\textnormal{T}$ and $Y_{n}^\textnormal{C}$ be the \textit{potential outcomes} \cite{Rubin74} of subject $n$ that would occur under treatment and control conditions, respectively.
A potential outcome is one of possible two outcomes: one under treatment and another under control conditions.
In medicine, for example, a subject is a patient, an outcome is recovery from disease, and treatment is to take a specific drug.
Let $Z_n$ be the indicator of treatment ($Z_n=1$ if treated and $Z_n=0$ if not treated).
Observed outcome is expressed as: $Y_n = Z_n Y_{n}^\textnormal{T} + (1-Z_n)Y_{n}^\textnormal{C}$.
Note that $Y_n = Y_{n}^\textnormal{T}$ if $Z_n=1$ and $Y_n = Y_{n}^\textnormal{C}$ if $Z_n=0$.
The causal effect $\tau_n$ is defined as the difference between the potential outcomes: $\tau_n = Y^\textnormal{T}_n - Y^\textnormal{C}_n$.
However, $\tau_n$ can not be obtained since either $Y_{n}^\textnormal{T}$ or $Y_{n}^\textnormal{C}$ is observed for each subject.
Matching estimator \cite{Stuart10} estimates unobserved potential outcomes from the observed outcomes of the closest subjects.
\begin{equation}
\hat{Y}_{n}^\textnormal{T} = \frac{1}{|\mathcal{M}^\textnormal{T}(n)|} \sum_{m \in \mathcal{M}^\textnormal{T}(n)} Z_m Y_{m}, \quad
\hat{Y}_{n}^\textnormal{C} = \frac{1}{|\mathcal{M}^\textnormal{C}(n)|} \sum_{m \in \mathcal{M}^\textnormal{C}(n)} (1-Z_m) Y_{m},
\end{equation}
where $\mathcal{M}^\textnormal{T}(n)$ and $\mathcal{M}^\textnormal{C}(n)$ are sets of matched subjects under treatment and control conditions.
Matched samples are typically chosen by similarity of subjects' covariates, e.g., demographics or previous medical histories.
The causal effect $\tau_n$ is estimated as,
\begin{align}
\label{eq:tau_matching}
\hat{\tau}_{n} 
&= Z_n \left( Y_n - \hat{Y}_{n}^\textnormal{C} \right)
+ (1-Z_n) \left(\hat{Y}_{n}^\textnormal{T} - Y_n \right)\nonumber\\
&= Z_n \left( Y_n^\textnormal{T} - \hat{Y}_{n}^\textnormal{C} \right)
+ (1-Z_n) \left(\hat{Y}_{n}^\textnormal{T} - Y_n^\textnormal{C} \right).
\end{align}

In the field of causal inference, we are mostly interested in the average treatment effect (ATE) or the average treatment effect on the treated (ATT), hence we take average of the above estimate over the set of subjects $\mathcal{S}$ or the set of treated subjects $\mathcal{S}^\textnormal{T}$:
\begin{equation}
\bar{\tau}_{\textnormal{ATE}} = \frac{1}{|\mathcal{S}|} \sum_{n \in \mathcal{S}} \hat{\tau}_{n}, \quad
\bar{\tau}_{\textnormal{ATT}} = \frac{1}{|\mathcal{S}^\textnormal{T}|} \sum_{n \in \mathcal{S}^\textnormal{T}} \hat{\tau}_{n}.
\end{equation}
The higher the value, the treatment is considered to be more effective.

\subsection{Neighborhood Method for Recommender System}
Neighborhood methods are divided into user-based neighborhood (UBN) and item-based neighborhood (IBN) methods.
Let $\mathcal{U}$ and $\mathcal{I}$ be sets of users and items, respectively, and $u,v \in \mathcal{U}$ and $i,j \in \mathcal{I}$.
The predictions of UBN and IBN are expressed as follows.
\begin{equation}
\hat{Y}_{ui}^{\textnormal{UBN}} = \frac{\sum_{v \in \mathcal{N} (u)} w_{uv} Y_{vi}}{\sum_{v \in \mathcal{N} (u)} w_{uv}}, \quad
\hat{Y}_{ui}^{\textnormal{IBN}} = \frac{\sum_{j \in \mathcal{N} (i)} w_{ij} Y_{uj}}{\sum_{j \in \mathcal{N} (i)} w_{ij}},
\end{equation}
where $\mathcal{N} (u)$ and $\mathcal{N} (i)$ are the sets of neighborhood users for $u$ and neighborhood items for $i$, respectively.
The weights $w_{uv}$ and $w_{ij}$ depend on the similarity between user pairs $u$ and $v$, and between item pairs $i$ and $j$, respectively.

The similarities are calculated based on previous interactions.
In UBN, if user $u$ and user $v$ have positive interactions for same items, they are regarded to be similar.
Popular choices for the similarity measure include cosine similarity, Pearson correlation, and Jaccard index among others~\cite{Ning15}.
The cosine similarity between users is expressed as, $\cos(u,v) =\boldsymbol{Y}_{u*} \cdot \boldsymbol{Y}_{v*}/||\boldsymbol{Y}_{u*}|| ||\boldsymbol{Y}_{v*}||$,
where $\boldsymbol{Y}_{u*} \equiv [Y_{u1}, Y_{u2}, ...Y_{u|\mathcal{I}|}]$ and $\boldsymbol{Y}_{v*} \equiv [Y_{v1}, Y_{v2}, ...Y_{v|\mathcal{I}|}]$ are vectors representing previous interactions for $u$ and $v$, respectively.
Top $k$ users by the similarity measure are chosen as neighborhood $\mathcal{N} (u)$.
The weight $w_{uv}$ becomes $w_{uv} = (\cos(u,v))^\alpha$, where $\alpha$ is a scaling factor.
$\mathcal{N} (i)$ and $w_{ij}$ for IBN are derived analogously.

\section{Causality-Aware Neighborhood Method}
Using notations similar to Subsection \ref{subsec:matching_estimator}, the causal effect of recommending item $i$ to user $u$ is expressed as $\tau_{ui} = Y^\textnormal{T}_{ui} - Y^\textnormal{C}_{ui}$.
In this setting, treatments are recommendations ($Z_{ui}=1$ if recommended) and outcomes are users' interactions ($Y_{ui}=1$ means positive interactions, such as purchases).
Total interactions from recommendations is the sum of $\tau_{ui}$ in recommendation lists.
Hence, we want to estimate $\tau_{ui}$ and rank items by the estimates.
In this section, we unify the matching estimator in causal inference and the neighborhood methods for recommender systems, and propose causality-aware neighborhood methods to rank items for the causal effect of recommendations.

Estimating the unobserved potential outcomes is a key component for estimating the causal effect.
We can apply UBN or IBN for the estimates.
\begin{equation}
\label{eq:CUBN_woM}
\textnormal{UBN:} \;
\hat{Y}_{ui}^\textnormal{T} = \frac{\sum_{v \in \mathcal{N} (u)} w_{uv} Z_{vi}Y_{vi}}{\sum_{v \in \mathcal{N} (u)} w_{uv}Z_{vi}}, \;
\hat{Y}_{ui}^\textnormal{C} = \frac{\sum_{v \in \mathcal{N} (u)} w_{uv} (1-Z_{vi})Y_{vi}}{\sum_{v \in \mathcal{N} (u)} w_{uv}(1-Z_{vi})},
\end{equation}
\begin{equation}
\label{eq:CIBN_woM}
\textnormal{IBN:} \;
\hat{Y}_{ui}^\textnormal{T} = \frac{\sum_{j \in \mathcal{N} (i)} w_{ij} Z_{uj} Y_{uj}}{\sum_{j \in \mathcal{N} (i)} w_{ij}Z_{uj}}, \;
\hat{Y}_{ui}^\textnormal{C} = \frac{\sum_{j \in \mathcal{N} (i)} w_{ij} (1-Z_{uj}) Y_{uj}}{\sum_{j \in \mathcal{N} (i)} w_{ij}(1-Z_{uj})}.
\end{equation}
Note that these estimates require only observed variables.
Direct application of the matching estimator to our setting yields the formula below,
\begin{equation}
	\label{eq:tau_woM}
\hat{\tau}_{ui} 
= Z_{ui} \left( Y_{ui} - \hat{Y}_{ui}^\textnormal{C} \right)
+ (1-Z_{ui}) \left(\hat{Y}_{ui}^\textnormal{T} - Y_{ui} \right).
\end{equation}
The observed outcome $Y_{ui}$ is used either as $Y_{ui}^\textnormal{T}$ or $Y_{ui}^\textnormal{C}$.

However, user behavior is not deterministic and the observed outcome has a random noise.\footnote{If we focus on ATE or ATT, as often the case in causal inference, the random noise is not a severe problem since it disappears by taking average of large samples.
It becomes a problem when we want to rank items by the estimates for each item.}
Hence we mix the own interaction $Y_{ui}$ and the neighbor interactions $Y_{vi}$ or $Y_{uj}$ to reduce random noises.
More specifically, we include $u$ and $i$ in $\mathcal{N} (u)$ and $\mathcal{N} (i)$, respectively, and we set $w_{uu}=1$ and $w_{ii}=1$.

To further reduce the variance, we force the estimates to shrink to zero if they rely on a few neighbors with low similarity.
We introduce shrinkage parameters $\beta^\textnormal{T}$ and $\beta^\textnormal{C}$ for the estimates of $\hat{Y}_{ui}^\textnormal{T}$ and $\hat{Y}_{ui}^\textnormal{C}$, respectively, and add them in the denominator.
\begin{equation}
\label{eq:CUBN}
\textnormal{UBN:} \;
\hat{Y}_{ui}^\textnormal{T} = \frac{\sum_{v \in \mathcal{N'} (u)} w_{uv} Z_{vi}Y_{vi}}{\beta^\textnormal{T} +\sum_{v \in \mathcal{N'} (u)} w_{uv}Z_{vi}}, \;
\hat{Y}_{ui}^\textnormal{C} = \frac{\sum_{v \in \mathcal{N'} (u)} w_{uv} (1-Z_{vi})Y_{vi}}{\beta^\textnormal{C} + \sum_{v \in \mathcal{N'} (u)} w_{uv}(1-Z_{vi})},
\end{equation}
\begin{equation}
\label{eq:CIBN}
\textnormal{IBN:} \;
\hat{Y}_{ui}^\textnormal{T} = \frac{\sum_{j \in \mathcal{N'} (i)} w_{ij} Z_{uj} Y_{uj}}{\beta^\textnormal{T} + \sum_{j \in \mathcal{N'} (i)} w_{ij}Z_{uj}}, \;
\hat{Y}_{ui}^\textnormal{C} = \frac{\sum_{j \in \mathcal{N'} (i)} w_{ij} (1-Z_{uj}) Y_{uj}}{\beta^\textnormal{C} + \sum_{j \in \mathcal{N'} (i)} w_{ij}(1-Z_{uj})}.
\end{equation}
Here the sets of neighbors $\mathcal{N'} (u)$ and $\mathcal{N'} (i)$ include $u$ and $i$ themselves.
With Eqs. (\ref{eq:CUBN}) and (\ref{eq:CIBN}), we estimate the causal effect as,
\begin{equation}
	\label{eq:tau}
\hat{\tau}_{ui} 
=\hat{Y}_{ui}^\textnormal{T} - \hat{Y}_{ui}^\textnormal{C},
\end{equation}
where the own interaction $Y_{ui}$ is included in either $\hat{Y}_{ui}^\textnormal{T}$ or $\hat{Y}_{ui}^\textnormal{C}$ depending on $Z_{ui}$.
Finally, to generate recommendation lists, items are ranked by the descending order of $\hat{\tau}_{ui}$ for each user.

We call our causality-aware user-based and item-based neighborhood methods as CUBN and CIBN, respectively.
To calculate similarity of users or items, we can use previous interactions, similar to original UBN and IBN.
We can also use the similarity based on previous treatment assignments $\boldsymbol{Z}_{u*} \equiv [Z_{u1}, Z_{u2}, ...Z_{u|\mathcal{I}|}]$ since we can expect that similar users receive similar recommendations if recommendations are properly personalized.
We suffix -O or -T in the names of our methods to clarify whether outcomes or treatment assignments are used.
The pseudo code of CUBN-O is shown in Algorithm \ref{algo:CUBN}.
Here $\textnormal{rank}_{u} (\hat{\tau}_{ui})$ is the ranking position of item $i$ for user $u$ when items are sorted by $\hat{\tau}_{ui}$ in descending order.
Cosine similarity is used in this work.
To obtain the algorithm for CUBN-T, line 4 is substituted with $w_{uv} \gets (\boldsymbol{Z}_{u*} \cdot \boldsymbol{Z}_{v*}/||\boldsymbol{Z}_{u*}|| ||\boldsymbol{Z}_{v*}||)^\alpha$.

\begin{algorithm}[tp] 
	\caption{Causality-aware User-Based Neighborhood method by Outcome similarity ({\textit{CUBN-O}}).}
	\label{algo:CUBN}
	\DontPrintSemicolon
	\KwIn{$k$, $\alpha$, $\beta^\textnormal{T}$, $\beta^\textnormal{C}$, $\{Y_{ui}\}$, $\{Z_{ui}\}$}
	\KwOut{$\{L_u |u \in \mathcal{U}\}$}
		\tcp{Phase1: neighborhood preparation }
	\For{$u \in \mathcal{U}$}{ 
		\For{$v \in \mathcal{U}$}{
			$w_{uv} \gets \left(\frac{\boldsymbol{Y}_{u*} \cdot \boldsymbol{Y}_{v*}}{||\boldsymbol{Y}_{u*}|| ||\boldsymbol{Y}_{v*}||} \right)^\alpha$
			\tcp*[r]{cosine similarity with scaling}
		}	
		$\mathcal{N'}(u) \gets \arg \max_{\mathcal{G}(u) \subset \mathcal{U},|\mathcal{G}(u)|=k } 
		\sum_{v \in \mathcal{G}(u)} w_{uv} $
		\tcp*[r]{top-$k$ neighbors}
	}
		\tcp{Phase2: item ranking}
	\For{$u \in \mathcal{U}$}{ 
		\For{$i \in \mathcal{I}$}{
			$\hat{\tau}_{ui} \gets 
			\frac{\sum_{v \in \mathcal{N'} (u)} w_{uv} Z_{vi}Y_{vi}}{\beta^\textnormal{T}  +\sum_{v \in \mathcal{N'} (u)} w_{uv}Z_{vi}} -
			\frac{\sum_{v \in \mathcal{N'} (u)} w_{uv} (1-Z_{vi})Y_{vi}}{\beta^\textnormal{C}  + \sum_{v \in \mathcal{N'} (u)} w_{uv}(1-Z_{vi})}$
		}	
		$L_u \gets \{ \textnormal{rank}_{u} (\hat{\tau}_{ui}) |i \in \mathcal{I}\}$
		\tcp*[r]{ranking list by descending order of $\hat{\tau}_{ui}$}
	}
	\Return{$\{L_u |u \in \mathcal{U}\}$}
\end{algorithm}

Standard collaborative filtering methods use only interaction logs $\{Y_{ui}\}$.
Our methods require previous recommendation logs $\{Z_{ui}\}$ in addition.
We assume that a certain recommender is already deployed in the service and we have the logs of the recommender.\footnote{Note that the deployed recommender is different from recommenders that we train and evaluate from $\{Y_{ui}\}$ and $\{Z_{ui}\}$, hence we might not have control over previous recommendation logs. In experiment section, we also investigate how different conditions of previous recommendations affect the proposed recommenders.}
Recommendation logs are commonly needed for previous methods targeting the causal effect~\cite{Bodapati08,Sato16,Sato19,Sato20}.
The previous IPS-based method~\cite{Sato20} further requires propensity, i.e., the probability of recommendations.
Our methods do not use propensity, hence we believe they are easier to deploy.

Our methods are based on standard assumptions of causal inference: ignorability, no interference, and no multiple versions~\cite{Imbens15,Hernan20}.\footnote{The latter two taken together are called the \textit{stable unit treatment value assumption} (SUTVA).}
The \textit{ignorability} assumption implies that treatment assignment ($Z_{ui}$) is independent of the potential outcomes ($Y_{ui}^\textnormal{T}, Y_{ui}^\textnormal{C}$) given the covariates ($X_u, X_i$): $Y_{ui}^\textnormal{T}, Y_{ui}^\textnormal{C} \perp Z_{ui} | X_u, X_i$ (see also causal graph of Fig. 1 (b) in~\cite{Sato20}).
Here $X_u$ and $X_i$ are features of user $u$ and item $i$, respectively.
We assume that user neighbors $\mathcal{N} (u)$ and item neighbors $\mathcal{N} (i)$ have features similar to user $u$ and item $i$, respectively.
The \textit{no interference} assumption means that a recommendation ($Z_{ui}$) does not affect other users' or items' outcomes ($Y_{vi}$ or $Y_{uj}$).
As a result of this assumption, there is no influence by item sequences in recommendation lists.
The \textit{no multiple versions} assumption states that there is only a single version of recommendation.
There could be several ways to recommend items, such as browser pop-ups and sending e-mails, but we assume that only one way is chosen for each dataset.
Relaxing these assumptions is an active area of research in causal inference~\cite{Hudgens08,Wang19b,Tyler13} and is also interesting future direction of this study.

\section{Experiments}
\subsection{Experimental Settings\protect\footnote{The codes and chosen hyper parameters for each method are available as ancillary files at  http://arxiv.org/abs/2012.09442.}}

\subsubsection{Datasets}
We used the MovieLens (ML)\footnote{https://grouplens.org/datasets/movielens} 100K and 1M datasets, and the Dunnhumby (DH)\footnote{https://www.dunnhumby.com/careers/engineering/sourcefiles} dataset.
The ML datasets \cite{Harper15} contains five-star movie ratings.
The DH dataset contains purchase and promotion logs from grocery stores.
For DH, we followed procedure described in \cite{Sato20} to generate a semi-synthetic dataset in \textit{Original} (DH-Ori) and \textit{Personalized} (DH-Per) settings.
For ML, we generated semi-synthetic datasets as follows,
\begin{enumerate}
	\item The ratings of all user-item pairs $\{\hat{R}_{ui}\}$ were predicted using rating matrix factorization~\cite{Koren09}.
	
	\item The probabilities of observing the ratings $\{\hat{O}_{ui}\}$ were predicted using logistic matrix factorization~\cite{Johnson14}.
	
	\item The probabilities of positive outcomes with and without recommendations were formulated as follows.
	\begin{equation}
	\label{eq:prob_prior}
	\mu_{ui}^\textnormal{T} = \sigma (\hat{R}_{ui} - \epsilon), \quad 
	\mu_{ui}^\textnormal{C} = \hat{O}_{ui}.
	\end{equation}
	Here $\sigma$ is a sigmoid function that converts predicted ratings $\hat{R}_{ui} \in [1,5]$ to probabilities $\mu_{ui}^\textnormal{T} \in [0,1]$.
	We set $\epsilon = 5.0$ the same as \cite{Saito20a}.
	
	\item The propensities were determined by users' preferences to items.
	\begin{equation}
	\label{eq:gen_propensity}
	P_{ui} = \min \left(1, a \left( 1/\textnormal{rank}_u \right)^b \right).
	\end{equation}
	Here $\textnormal{rank}_u$ is item rankings by $\mu_{ui}^\textnormal{T}+ \mu_{ui}^\textnormal{C}$.
	The parameters $a$ and $b$ control the average and the unevenness of propensities, respectively.
	We set $b=1.0$ for the default condition.
	The average number of recommendations for users was set to 100 by adjusting $a$.
	
	\item The potential outcomes under treatment and control conditions, and recommendation assignments were sampled as follows.
	\begin{equation}
	Y_{ui}^\textnormal{T} \sim \textnormal{Bernoulli} (\mu_{ui}^\textnormal{T}), \quad
	Y_{ui}^\textnormal{C} \sim \textnormal{Bernoulli} (\mu_{ui}^\textnormal{C}), \quad
	Z_{ui} \sim \textnormal{Bernoulli} (P_{ui}).
	\end{equation}
	Then, causal effect $\tau_{ui}$ and observed outcome $Y_{ui}$ were obtained as,
	\begin{equation}
	\tau_{ui} = Y_{ui}^\textnormal{T} - Y_{ui}^\textnormal{C}, \quad
	Y_{ui} = Z_{ui} Y_{ui}^\textnormal{T} +(1-Z_{ui}) Y_{ui}^\textnormal{C}.
	\end{equation}
	Note that $\tau_{ui}$ was provided only for evaluation.
	This sampling can be repeated $n$ times for each user-item pair.
	We independently sampled training, validation, and test data, and used for the purposes.
\end{enumerate}

The steps 1, 2 and 3 are similar to that of \cite{Saito20a}.
The steps 4 and 5 are similar to steps 3 and 4 of \cite{Sato20}.
Unlike~\cite{Sato20}, we generated only one observation for each user-item pair for training data (i.e., we set $n_{train}=1$ as opposed to $n_{train}=10$ in \cite{Sato20}) since this setting more directly reflects the unobservable nature of the causal effect.
The reasoning of Eq. (\ref{eq:prob_prior}) in step 3 is as follows.
A choice of a movie to watch (${O}_{ui}$) may be said to depend on expected entertainment from watching it.
A rating (${R}_{ui}$) reflects the experienced entertainment value after watching the movie.
If a user knew the entertainment value before consumption, the user would choose movies based on this.
Recommendations are often provided with explanations~\cite{Tintarev2015} and the explanations  help users predict entertainment values of items~\cite{Bilgic05,Tintarev08}.
Hence we related the watching probability with recommendation $\mu_{ui}^\textnormal{T}$ to experienced entertainment value ${R}_{ui}$, and the watching probability without recommendation $\mu_{ui}^\textnormal{C}$ to users' natural watching behavior ${O}_{ui}$.

The statistics of generated datasets are summarized in Table \ref{tab:stat_data}.
ATE over whole user-item pairs are positive, meaning that recommendations generally tend to promote user interactions.
We also confirmed that $\mu_{ui}^\textnormal{T} > \mu_{ui}^\textnormal{C}$ for about 90\% of user-item pairs in the ML datasets and about 80\% of user-item pairs in the DH datasets.
However, $\mu_{ui}^\textnormal{T} < \mu_{ui}^\textnormal{C}$ for the remaining pairs and thus $\tau_{ui}$ tend to be negative for those pairs.
Recommendations can have negative impact when they create bad feelings for users, e.g., creepiness~\cite{Torkamaan19}.
Note that $\tau_{ui}$ can become negative by the randomness of user behaviors when $\mu_{ui}^\textnormal{T} \approx \mu_{ui}^\textnormal{C}$.

\begin{table*}[htbp]
	\caption{Statistics of generated datasets.}
	\label{tab:stat_data}
	\centering
	\setlength{\tabcolsep}{2mm} 
	\begin{tabular}{lrrrrr}
		\toprule
		Dataset & \#User & \#Item & $\{Y_{ui}=1\}$ & $\{Z_{ui}=1\}$ & ATE  \\
		\midrule
		DH-Original & 2,309 & 1,372 & 35,010 & 483,660 & 0.0044 \\
		DH-Personalized & 2,309 & 1,372 & 37,731 & 483,727 & 0.0045 \\
		ML-100K & 943 & 1,682 & 92,523 & 94,054 & 0.0735\\
		ML-1M & 6,040 & 3,952 & 985,994 & 603,108 & 0.0981 \\
		\bottomrule
	\end{tabular}
\end{table*}

\newpage
\subsubsection{Compared Methods}
\label{subsubsec:compared_methods}
The following methods were compared.
\begin{itemize}
	\item \textbf{Random}: Items are ranked randomly.
	\item \textbf{Pop}: Items are ranked by popularity, i.e., number of positive outcomes.
	\item \textbf{UBN/IBN}: Traditional user-based and item-based neighborhood methods.
	\item \textbf{BPR}~\cite{Rendle09}: A commonly used pairwise learning method.
	\item \textbf{CausE}~\cite{Bonner18}: A joint training of prediction models for $Y_{ui}^\textnormal{T}$ and $Y_{ui}^\textnormal{C}$.
	\item \textbf{ULRMF/ULBPR}~\cite{Sato19}: Pointwise and pairwise learning methods for $\tau_{ui}$. 
	\item \textbf{DLTO/DLCE}~\cite{Sato20}: IPS-based unbiased learning methods for $Y_{ui}^\textnormal{T}$ and $\tau_{ui}$.
	\item \textbf{CUBN/CIBN}: Our causality-aware user-based and item-based neighborhood methods for $\tau_{ui}$.
\end{itemize}
By comparing CUBN/CIBN and UBN/IBN, we verify whether our methods successfully extend UBN/IBN for the causal effect.
We also compare our neighborhood methods with previous model-based methods targeting the causal effect: ULBPR, ULRMF, and DLCE.
Previous research~\cite{Sato19,Sato20} shows that CausE and DLTO are also strong baselines, hence we included them.
Our methods can use treatment assignments or positive outcomes for calculating user/item similarities.
We suffix -T or -O to clarify which one is used.
To investigate the effectiveness of mixing own and neighbor interactions, we also experimented on our methods without the mixture (-woM), i.e., Eqs. (\ref{eq:CUBN_woM})-(\ref{eq:tau_woM}) are used instead of Eqs. (\ref{eq:CUBN})-(\ref{eq:tau}).

\subsubsection{Evaluation Protocols}
Commonly used accuracy metrics, such as precision, reward positive interactions even if that would occur in the absence of recommendation (e.g., item A in Fig. 1.)
We want to reward positive interactions purely caused by recommendation (e.g., item B in Fig. 1), and the accuracy metrics is not suitable (see also Section 2.1 in \cite{Sato19}).
Hence, we used the causal variants of precision@n (CP@n), discounted cumulative gain (CDCG), and average rank (CAR) \cite{Sato20}.
They are expressed respectively as,
\begin{equation}
\textnormal{Causal Precision@\textit{n} (CP@\textit{n}): } \sum_i \frac{\boldsymbol{1}(\textnormal{rank}_u (\hat{s}_{ui}) \leq n) \tau_{ui}}{n},
\end{equation}
\begin{equation}
\textnormal{Causal DCG (CDCG): } \sum_i \frac{\tau_{ui}}{\log_2 (1+\textnormal{rank}_u (\hat{s}_{ui}))},
\end{equation}
\begin{equation}
\textnormal{Causal Average Rank (CAR): } \frac{1}{I} \sum_i \textnormal{rank}_u (\hat{s}_{ui}) \tau_{ui},
\end{equation}
where $\hat{s}_{ui}$ is the predicted score of item $i$ for user $u$ and $\textnormal{rank}_u (\hat{s}_{ui})$ is the ranking position of the item.
Items are ranked by the descending order of $\hat{s}_{ui}$.
In our methods, items are ranked by the causal effect estimates $\hat{\tau}_{ui}$, i.e., $\hat{s}_{ui}=\hat{\tau}_{ui}$.
We calculated the above metrics for each user and took average over all users.
Note that $\tau_{ui}$ is a ternary variable ($\tau_{ui} \in \{1,0,-1\}$) and the metrics can be negative.

The hyper parameters of each method were tuned with validation data to optimize each metric, i.e., chosen parameters were different for each metric.
We used the same shrinkage parameters for treatment and control ($\beta=\beta^\textnormal{T}=\beta^\textnormal{C}$).
The exploration ranges for the proposed methods were as follows: the maximum number of neighbors $\in \{10,30,100,300,1000,3000,10000\}$, the scaling factor $\alpha \in \{0.33,0.5,1.0,2.0,3.0,5.0\}$, and the shrinkage parameter $\beta \in \{0,0.3,1,3,10,30,100\}$.
The exploration ranges for other baselines were same with~\cite{Sato20}.

\subsection{Results and Discussions}

\subsubsection{Performance Comparison}

Tables \ref{tab:comparison_DH} and \ref{tab:comparison_ML} show the performance comparison.
The best among previous methods differ for datasets.
Our CUBNs constantly outperform them in all datasets.
CIBNs perform worse but are still competitive to other baselines.
CUBN-O and CUBN-T tend to perform similarly, and any differences depend on datasets and metrics.
CUBN-O uses previous outcomes for user similarities same as traditional UBN.
On the other hand, CUBN-T uses previous treatment assignments for user similarities that is original to our work.
The result indicates that similarity of previous treatment assignments can provide good measure of user similarities.
Furthermore, CUBN and CIBN counterparts not using own and neighborhood interaction mixtures (-woM) are often outperformed by methods which do, showing its importance.

\begin{table*}[htbp]
	\small
	\caption{Performance comparison in the Dunnhumby (DH) dataset. 
		The best results are highlighted in bold.
		Note that the smaller is better in CAR.
	}
	\label{tab:comparison_DH}
	\centering
	\scalebox{1.0}{
		\begin{tabular}{lcccccccc}
			\hline
			& \multicolumn{4}{c}{DH-Original} & \multicolumn{4}{c}{DH-Personalized}\\
			\cmidrule(lr){2-5} \cmidrule(lr){6-9} 
			& CP@10 & CP@100 & CDCG & CAR & CP@10 & CP@100 & CDCG & CAR    \\
			\hline
			Random & 0.0046 & 0.0049 & 0.726 & 3.01
			& 0.0048 & 0.0044 & 0.672  & 2.84 \\
			Pop & 0.0293 & 0.0157 & 0.925 & 1.86
			& 0.0275 & 0.0131 & 0.858 & 1.64 \\
			BPR & 0.0331 & 0.0153 & 0.923 & 1.86
			& 0.0564 & 0.0187 & 0.858 & 1.54 \\
			UBN & 0.0294 & 0.0153 & 0.926 & 1.87
			& 0.0419 & 0.0190 & 0.922 & 1.36 \\
			IBN & 0.0301 & 0.0138 & 0.903 & 1.94
			& 0.0438 & 0.0179 & 0.928 & 1.49  \\
			\hline
			CausE & 0.0337 & 0.0204 & 1.009 & 1.95
			& 0.0857 & 0.0186 & 1.110 & 1.39 \\
			ULRMF & 0.0359 & 0.0168 & 0.937 & \textbf{1.78}
			& 0.0802 & 0.0203 & 1.005  & 1.39\\
			ULBPR & 0.0343 & 0.0143 & 0.918 & 1.80
			& 0.0806 & 0.0209 & 1.038 & 1.32 \\
			DLTO & 0.0358 & 0.0151 & 0.955 & 1.82
			& 0.0813 & 0.0198 & 1.063 & 1.41 \\
			DLCE & 0.0354 & 0.0116 & 0.882 & 2.70
			& 0.0839 & 0.0209 & 1.036 & 1.38 \\
			\hline
			CUBN-O & 0.0424 & 0.0193 & 0.986 & 1.98
			& 0.0877 & 0.0240 & 1.124 & 1.24 \\
			CUBN-T & \textbf{0.0513} & \textbf{0.0216} & \textbf{1.030} & \textbf{1.78}
			& 0.0890 & \textbf{0.0257} & 1.112 & \textbf{1.13}\\
			CIBN-O & 0.0328 & 0.0110 & 0.892 & 2.43
			& 0.0871 & 0.0190 & 1.112 & 1.36 \\
			CIBN-T & 0.0301 & 0.0095 & 0.872 & 2.61
			& 0.0889 & 0.0181 & \textbf{1.135}  & 1.61\\
			\hline
			CUBN-O-woM  & 0.0437 & 0.0186 & 0.979 & 2.20
			& \textbf{0.0902} & 0.0199 & 1.107  & 1.30 \\
			CUBN-T-woM  & 0.0436 & 0.0198 & 0.991 & 2.10
			&0.0901 & 0.0124 & 1.005  & 2.40\\
			CIBN-O-woM  & 0.0382 & 0.0140 & 0.909 & 2.38
			& 0.0738 & 0.0175 & 1.008  & 1.39 \\
			CIBN-T-woM  & 0.0333 & 0.0098 & 0.890 & 2.69
			&0.0881 & 0.0168 & 1.098  & 2.03 \\
			\hline
		\end{tabular}
	}
\end{table*}

\begin{table*}[htbp]
	\small
	\caption{Performance comparison in the MovieLens (ML) 100K and 1M datasets. 
		The best results are highlighted in bold.
		Note that the smaller is better in CAR.
	}
	\label{tab:comparison_ML}
	\centering
	\scalebox{1.0}{
		\begin{tabular}{lcccccccc}
			\hline
			& \multicolumn{4}{c}{ML-100K} & \multicolumn{4}{c}{ML-1M}\\
			\cmidrule(lr){2-5} \cmidrule(lr){6-9}
			& CP@10 & CP@100 & CDCG & CAR  & CP@10 & CP@100 & CDCG & CAR  \\
			\hline
			Random 
			& 0.076 & 0.075 & 13.9 & 61.8
			& 0.097 & 0.098 & 38.0  & 194\\
			Pop 
			&-0.215 & -0.085 & 11.3 & 73.7
			&-0.135 & -0.042 & 35.5 & 196\\
			BPR 
			& 0.092 & 0.088 & 14.0 & 61.7
			& 0.102 & 0.103 & 38.1 & 194 \\
			UBN 
			&-0.217 & -0.102 & 11.1 & 66.6
			&-0.175 & -0.058 & 35.2 & 165 \\
			IBN 
			& 0.098 & 0.099 & 14.0 & 63.2
			& 0.052 & 0.055 & 36.8 & 177  \\
			\hline
			CausE 
			& 0.310 & 0.214 & 16.4 & 34.4
			& 0.309 & 0.246 & 42.4 & 122 \\
			ULRMF 
			& 0.302 & 0.148 & 15.8 & 39.0
			& 0.160 & 0.152 & 39.9  & 152\\
			ULBPR 
			& 0.333 & 0.163 & 15.6 & 43.9
			& 0.245 & 0.187 & 40.4 & 143 \\
			DLTO 
			& 0.330 & 0.155 & 15.3 & 53.2
			& 0.289 & 0.202 & 40.5 & 152 \\
			DLCE 
			& 0.330 & 0.215 & 16.6 & 28.8
			& 0.319 & \textbf{0.258} & 42.4 & 119 \\
			\hline
			CUBN-O 
			& 0.349 & \textbf{0.218} & \textbf{16.9} & 27.2
			& 0.334 & \textbf{0.258} & \textbf{42.7} & 116 \\
			CUBN-T 
			& \textbf{0.350} & \textbf{0.218} & 16.8 & \textbf{25.9}
			& \textbf{0.336} & 0.256 & 42.6 & 127\\
			CIBN-O 
			& 0.184 & 0.145 & 15.5 & 30.0
			& 0.236 & 0.186 & 41.1 & 120 \\
			CIBN-T 
			& 0.160 & 0.149 & 15.6 & 31.8
			& 0.188 & 0.173 & 40.9  & 122\\
			\hline
			CUBN-O-woM  
			&0.310 & 0.194 & 16.6 & 29.0
			&0.291 & 0.233 & 42.4 & 115\\
			CUBN-T-woM  
			&0.311 & 0.194 & 16.6 & 29.0
			&0.294 & 0.237 & 42.4  & \textbf{114}\\
			CIBN-O-woM  
			&0.147 & 0.123 & 15.1 & 34.6
			&0.216 & 0.183 & 40.6 & 117\\
			CIBN-T-woM  
			&0.118 & 0.126 & 15.2 & 34.5
			&0.160 & 0.168 & 40.8 & 123\\
			\hline
		\end{tabular}
	}
\end{table*}

\subsubsection{Dependence on Hyper Parameters}
\begin{figure}[htbp]
	\begin{center}
		\subfigure[DH-Ori.]{\includegraphics[width=0.242\textwidth]{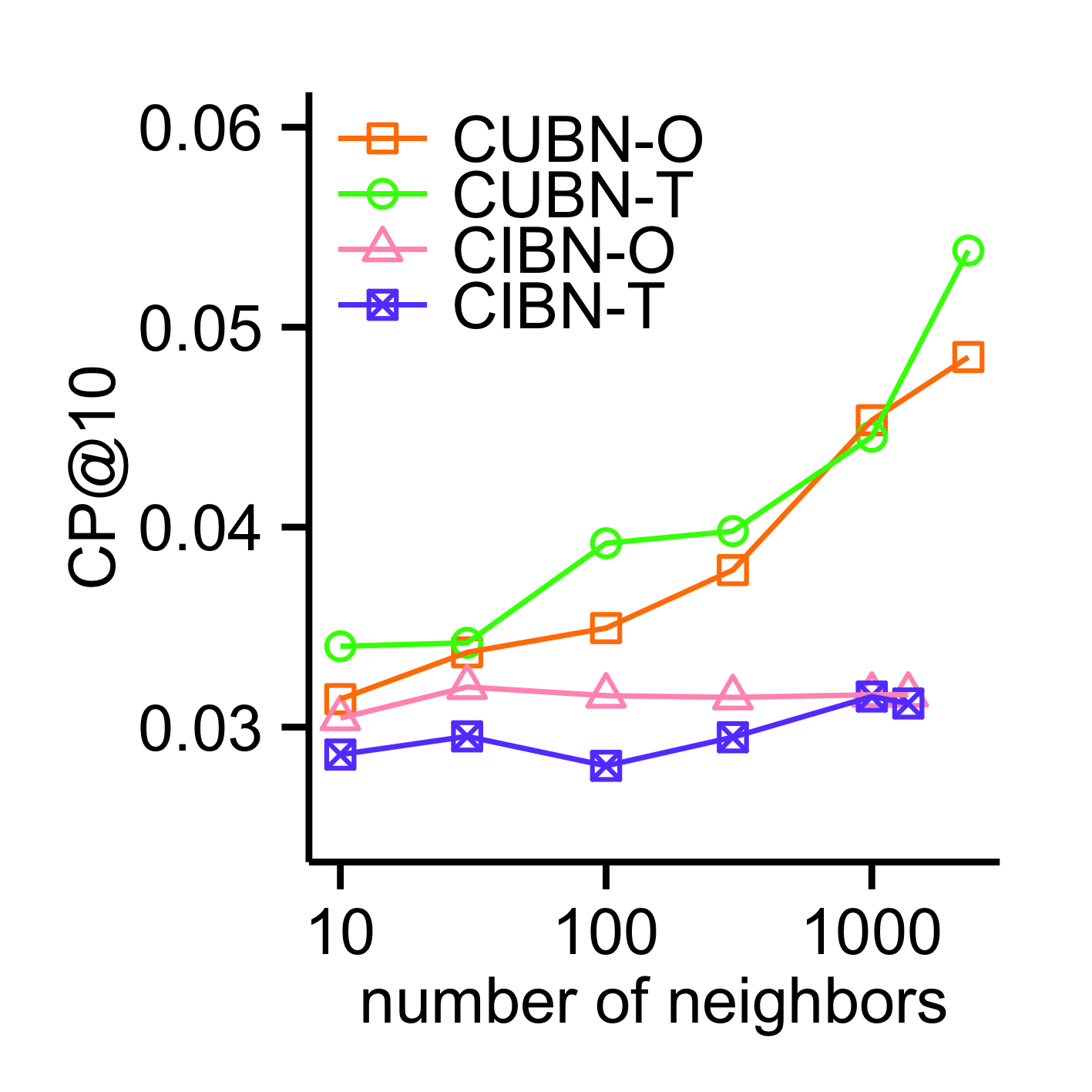}}
		\subfigure[DH-Per.]{\includegraphics[width=0.242\textwidth]{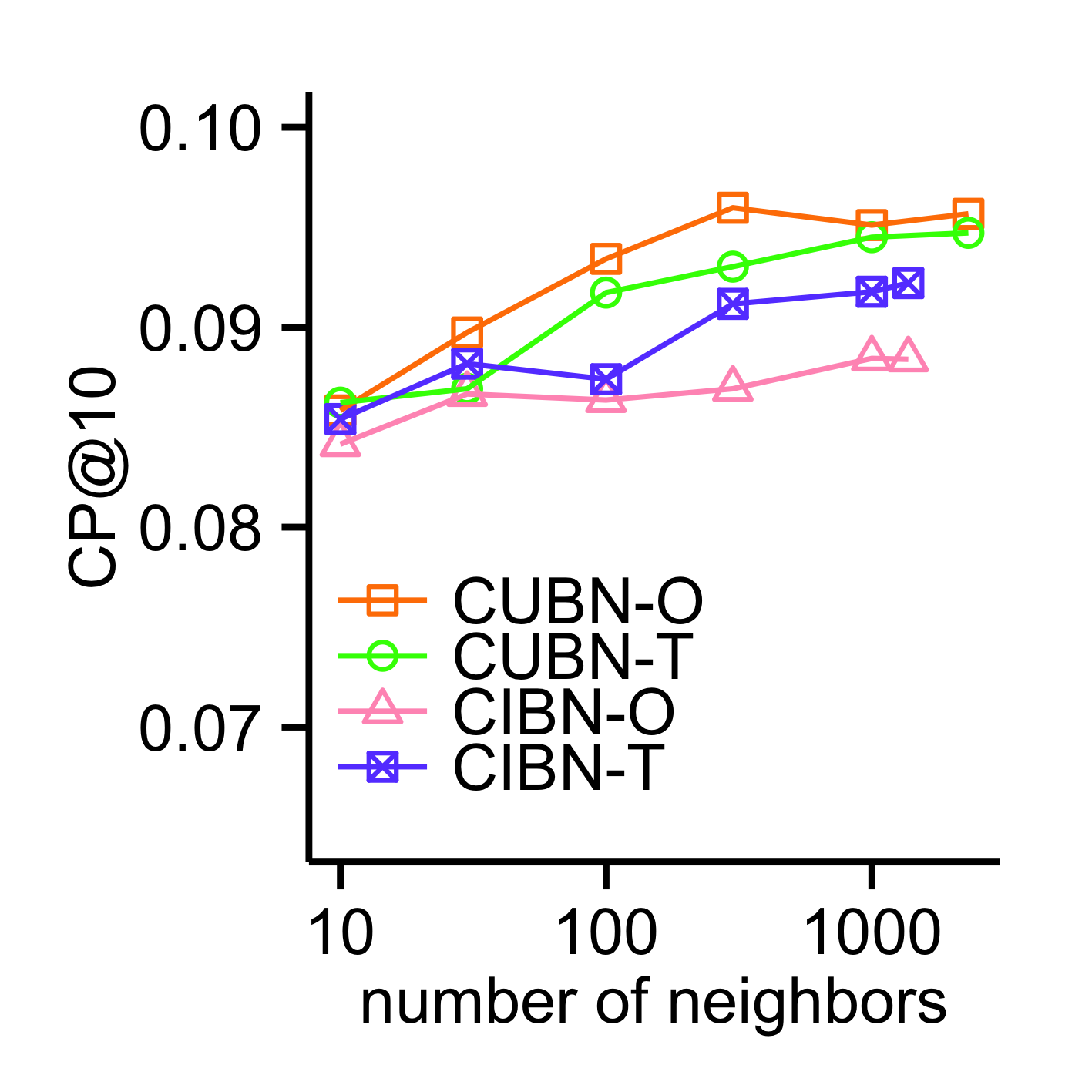}}
		\subfigure[ML-100K.]{\includegraphics[width=0.242\textwidth]{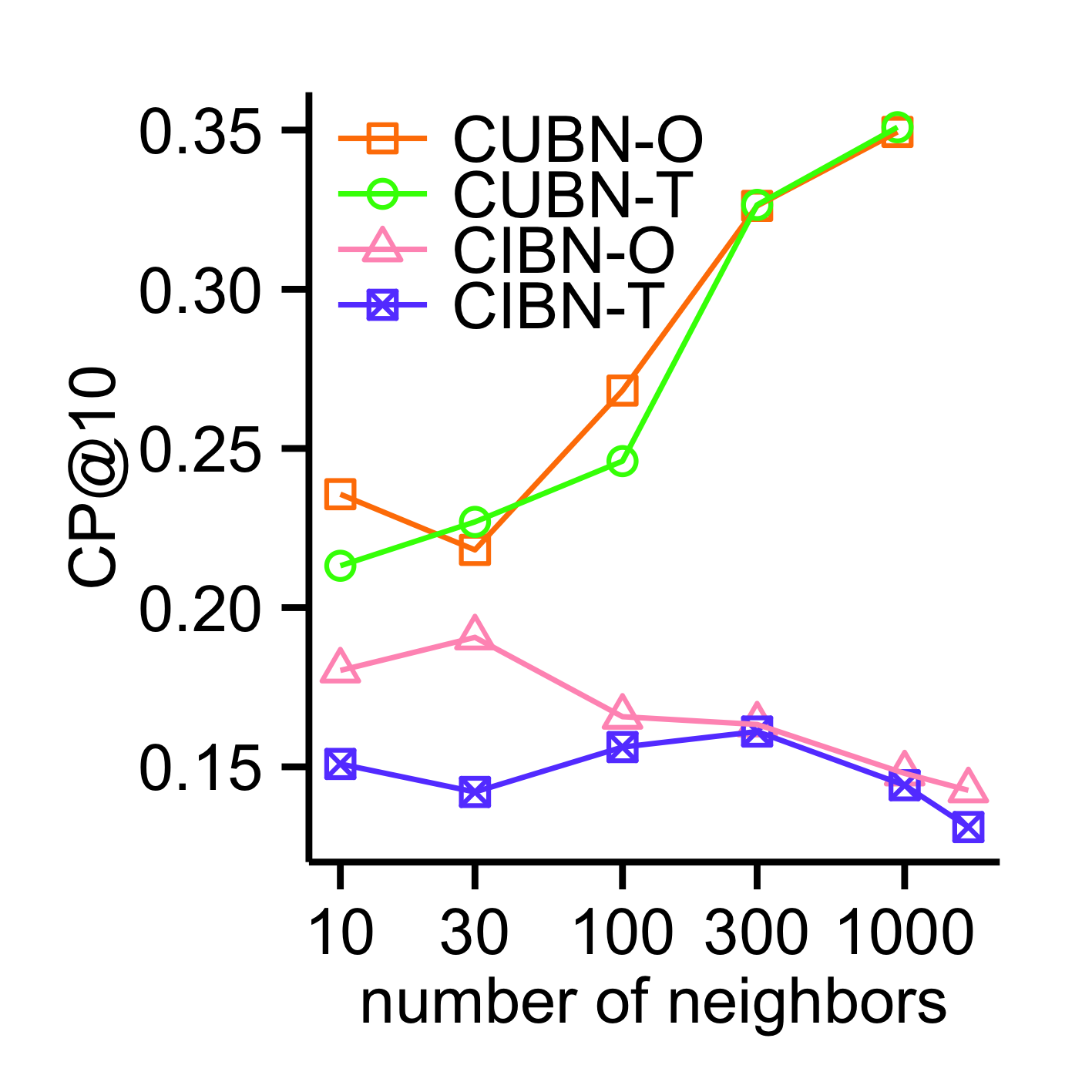}}
		\subfigure[ML-1M.]{\includegraphics[width=0.242\textwidth]{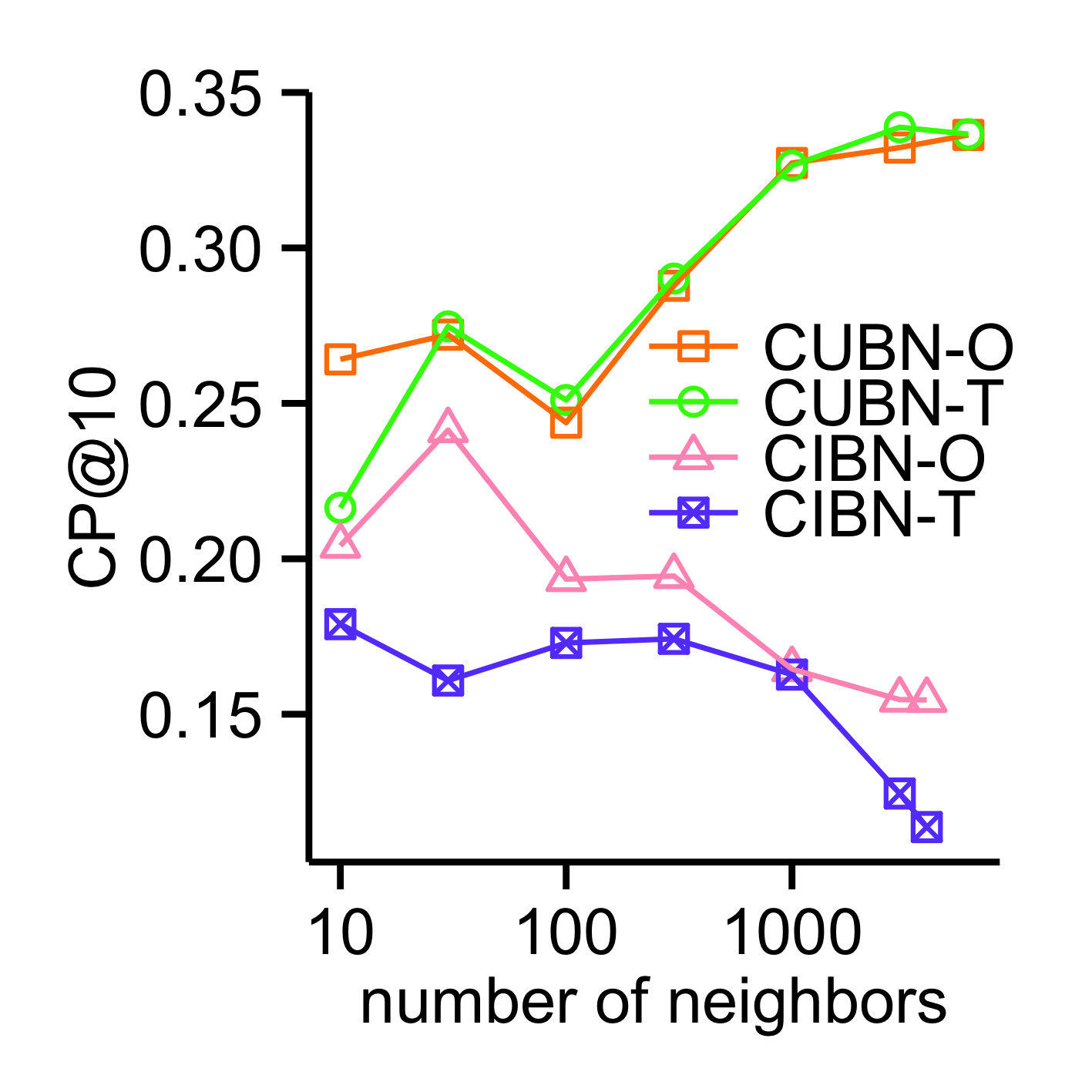}}
		\caption{Dependence on the number of neighbors in validation datasets. The scaling factor $\alpha$ and the shrinkage parameter $\beta$ are set to the optimal values for each number of neighbors. Note that the possible number of neighbors are restricted by either the number of users or that of items.}
		\label{fig:num_neighbor}
	\end{center}
\end{figure}

As our methods are neighborhood methods, the dependence on the number of neighbors is important.
Fig. \ref{fig:num_neighbor} shows the results.
General trends show that performance improves with increasing numbers of neighbors.
In ML-100K and ML-1M datasets, CIBNs reach maximum performance with relatively smaller numbers of neighbors.

\begin{figure}[ht]
	\begin{center}
		\subfigure[ML-1M (CP@10).]{\includegraphics[width=0.45\textwidth]{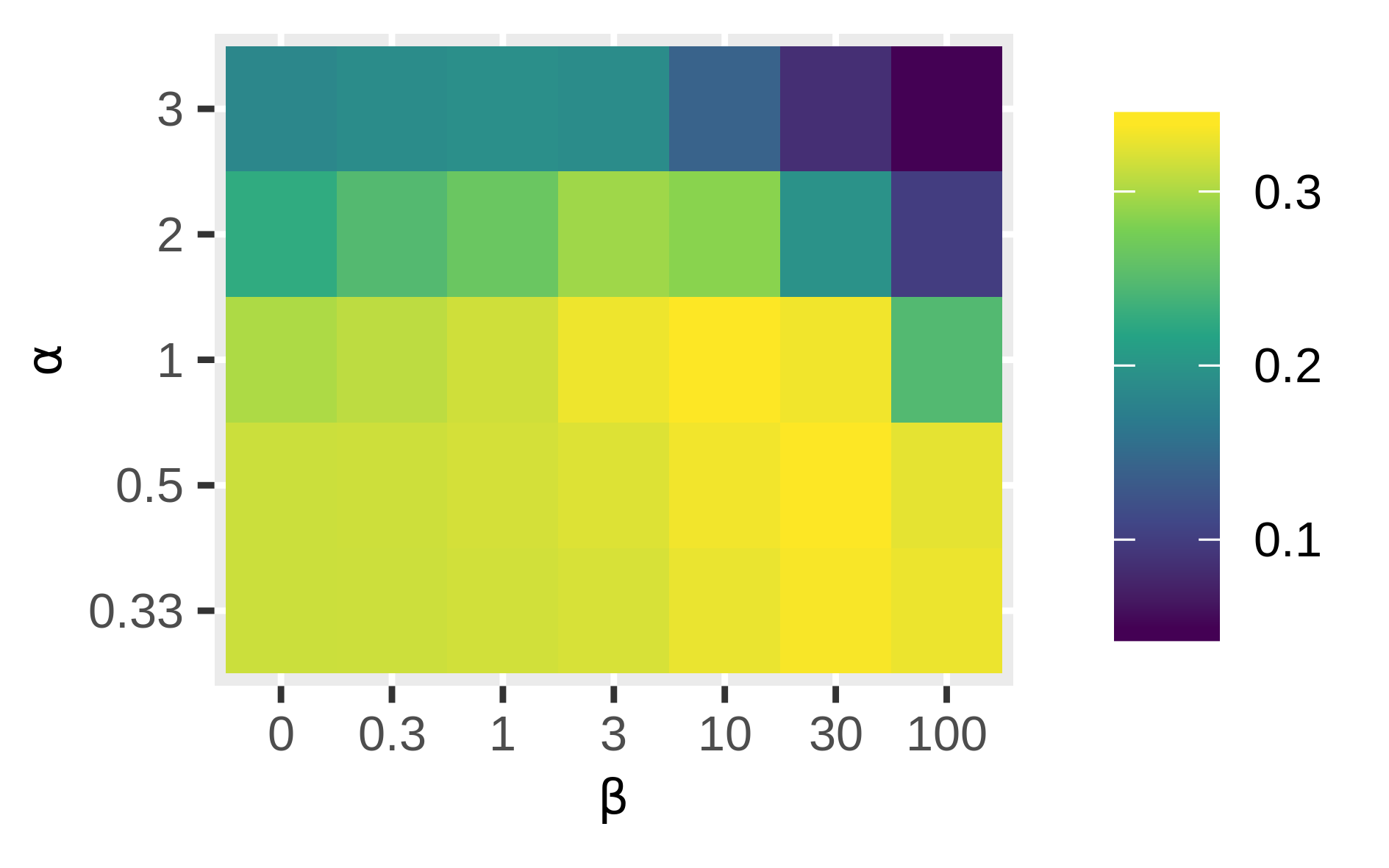}}
		\subfigure[ML-1M (CP@100).]{\includegraphics[width=0.45\textwidth]{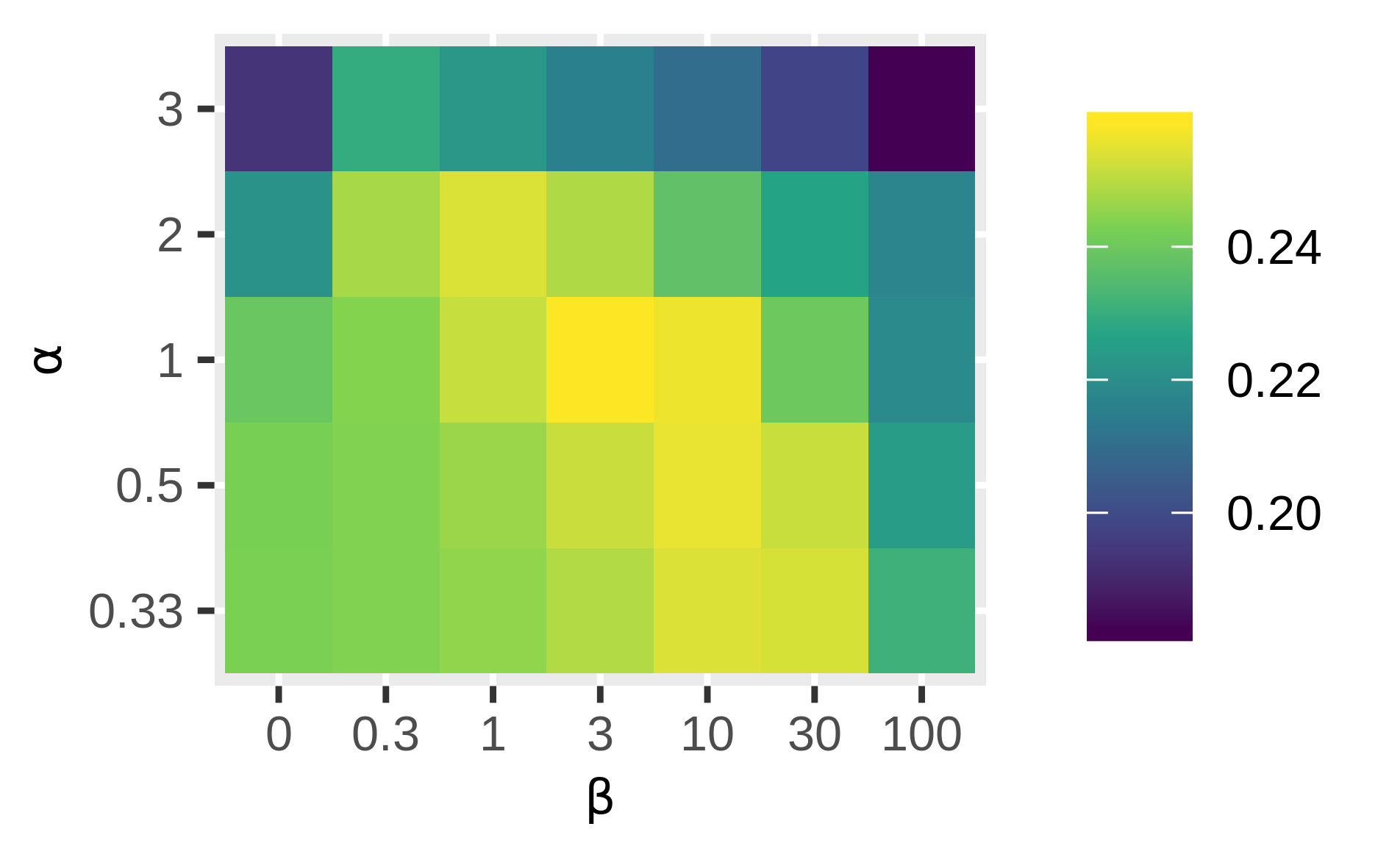}}
		\caption{Dependence on the scaling factor $\alpha$ and the shrinkage parameter $\beta$ for CUBN-O in the ML-1M dataset.
			The number of neighbors are set to 6,040.}
		\label{fig:alpha_beta}
	\end{center}
\end{figure}

Our methods have other two key hyper-parameters: the scaling factor $\alpha$ and the shrinkage parameter $\beta$.
We investigated the dependence on these parameters (Fig. \ref{fig:alpha_beta}).
The best performances were obtained at $\beta > 0$, showing the effectiveness of introducing the shrinkage.
Optimal $\beta$ for CP@10 is larger than that for CP@100.
This trend was similarly observed in other datasets.
We suppose that inappropriate item selection by random noise of causal effect estimates affects CP more severely when recommendation list is small, thus the shrinkage $\beta$ should be larger for CP@10.

\subsubsection{Influence of Difference in Previous Recommendation Logs}
\begin{figure}[htbp]
	\begin{center}
		\subfigure[CP@10.]{\includegraphics[width=0.242\textwidth]{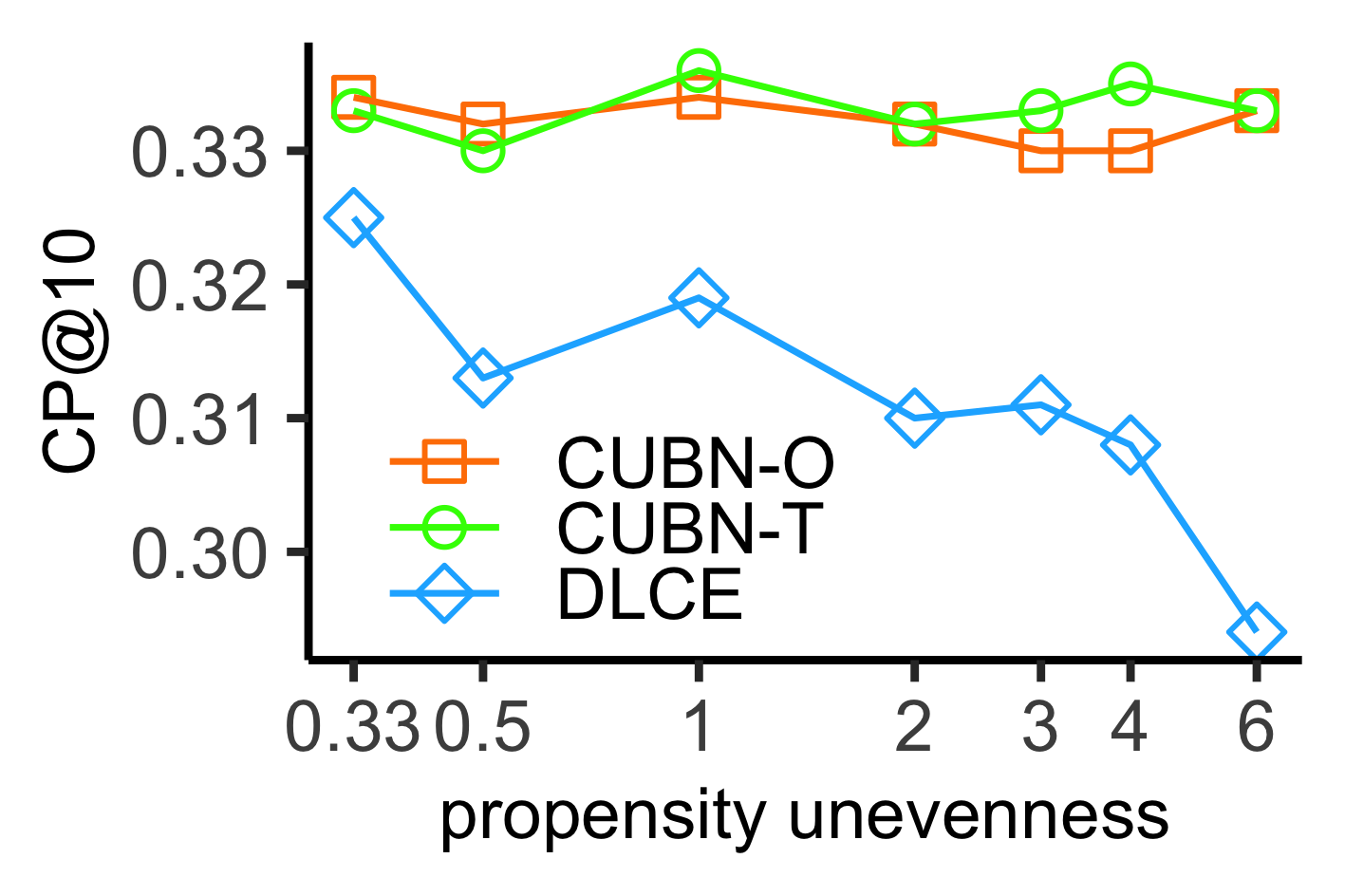}}
		\subfigure[CP@100.]{\includegraphics[width=0.242\textwidth]{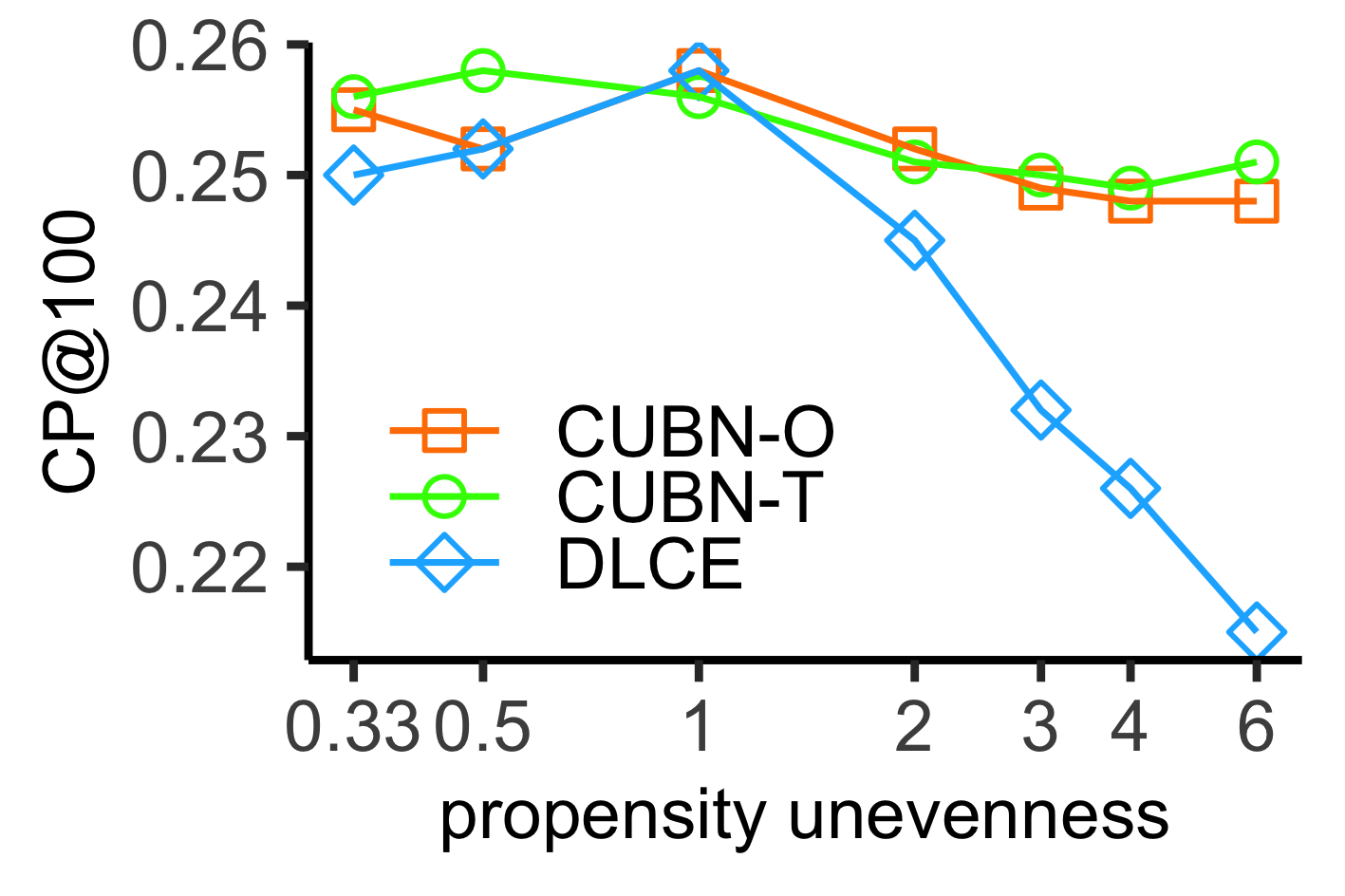}}
		\subfigure[CP@10.]{\includegraphics[width=0.242\textwidth]{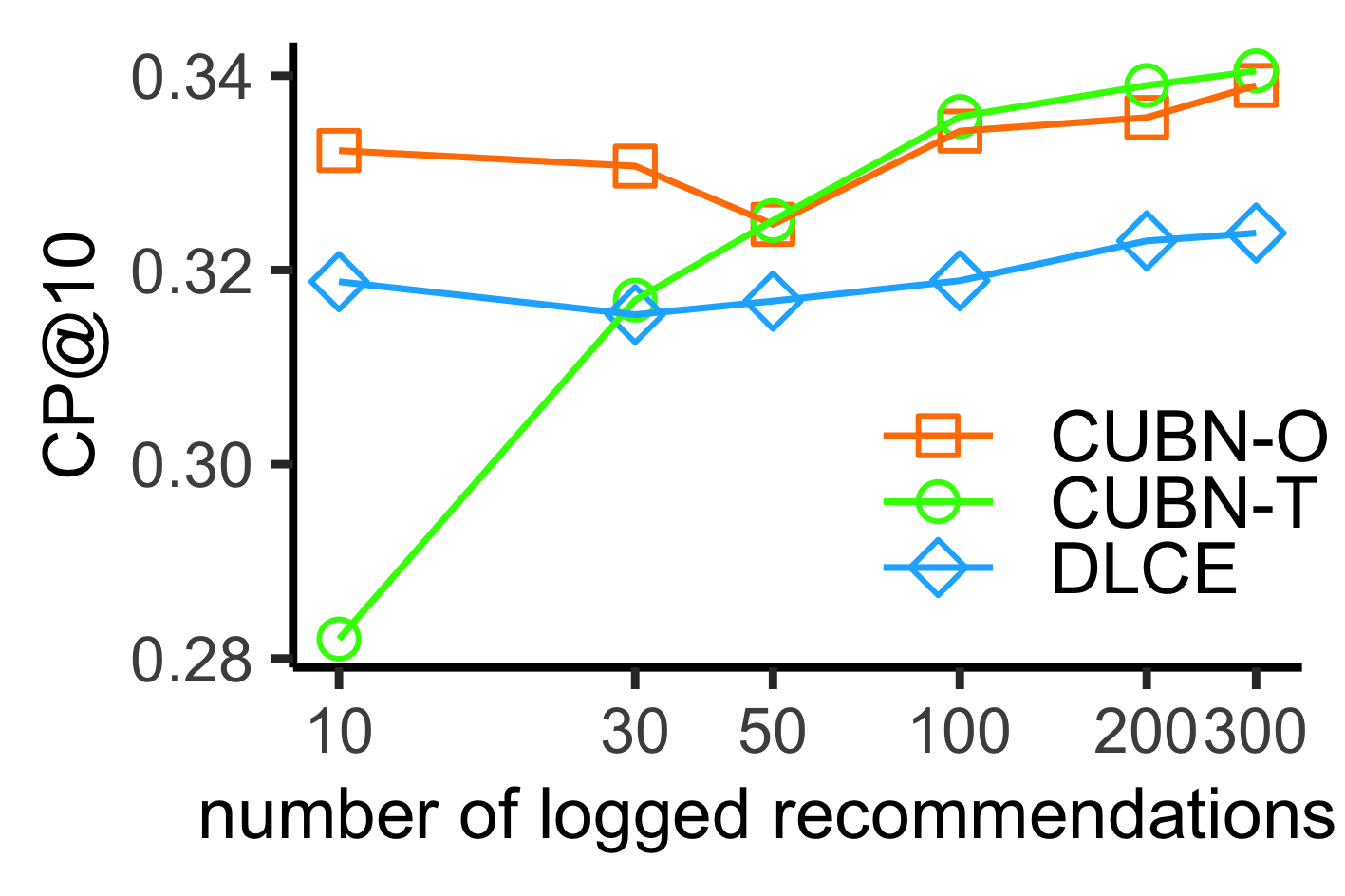}}
		\subfigure[CP@100.]{\includegraphics[width=0.242\textwidth]{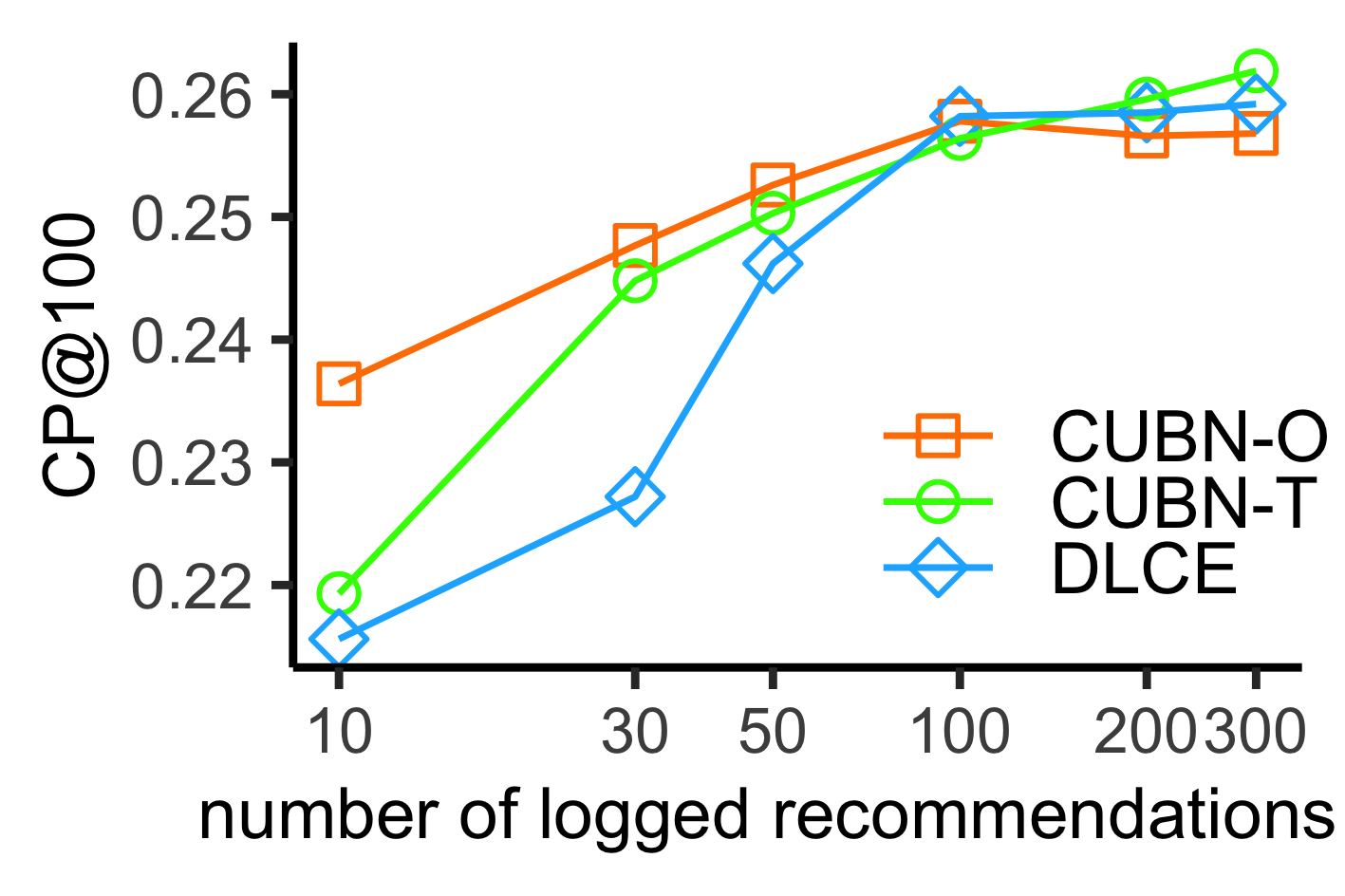}}
		\caption{Performances under the varied unevenness of propensity (a, b) and the varied number of logged recommendations per user (c, d).}
		\label{fig:unevenness_num_logged_rec}
	\end{center}
\end{figure}

IPS are known to suffer from variance by very small propensities.
This happens when recommendation assignments shift toward deterministic assignments, i.e., propensities are close to 0.0 or 1.0.
In our semi-synthetic data generation, increasing unevenness parameter $b$ in Eq. (\ref{eq:gen_propensity}) makes recommendations more deterministic.
Hence we investigated how it affects our methods and IPS-based previous method (DLCE).
As seen from Fig. \ref{fig:unevenness_num_logged_rec} (a, b), DLCE degrades with increasing unevenness.
On the other hand, our methods are more robust to this unevenness.

Recommendation methods targeting the causal effect commonly require recommendation logs.
Here we investigated how the number of logged recommendations for each user affects the performance.
For CP@10 (Fig. \ref{fig:unevenness_num_logged_rec} (c)), the performances of CUBN-O and DLCE are mostly stable, while CUBN-T degrades with less number of logged recommendations.
This is reasonable considering that CUBN-T obtains neighbors by the similarity of recommendation assignments.
For CP@100 (Fig. \ref{fig:unevenness_num_logged_rec} (d)), all methods are affected by the number of logged recommendations, but CUBN-O is relatively robust.

\section{Conclusions}
We proposed causality-aware neighborhood methods to generate item ranking by the causal effect of recommendations.
We unified traditional neighborhood-based recommendation methods with matching estimator, and further enhanced them by mixing the own and neighbor observations and introducing the shrinkage for potential outcome estimates.
Models proposed in this paper outperformed baselines on causal effect versions of commonly used ranking metrics. 
This was particularly true for models augmenting user-based neighborhood methods for causal effect. 
The results suggest that these models can lead to improved sales and user engagement and are thus highly beneficial for businesses employing recommender systems.
In the future work, our methods can be enhanced by applying graph-based neighborhood similarities~\cite{Fouss07,Luo08} or by learning neighborhood similarities~\cite{Ning11,Kabbur13}.
Another direction of future work is to leverage contextual information~\cite{Adomavicius05}.
Since neighborhood methods are known to be effective in session-based recommendations~\cite{Ludewig19}, it would be also interesting to extend our methods for session-based recommendations.

%
%
\bibliographystyle{splncs04}
\bibliography{arxiv}

\begin{thebibliography}{10}
\providecommand{\url}[1]{\texttt{#1}}
\providecommand{\urlprefix}{URL }
\providecommand{\doi}[1]{https://doi.org/#1}

\bibitem{Adomavicius05}
Adomavicius, G., Sankaranarayanan, R., Sen, S., Tuzhilin, A.: Incorporating
  contextual information in recommender systems using a multidimensional
  approach. ACM Trans. Inf. Syst.  \textbf{23}(1),  103–145 (Jan 2005).
  \doi{10.1145/1055709.1055714}, \url{https://doi.org/10.1145/1055709.1055714}

\bibitem{Agarwal19}
Agarwal, A., Takatsu, K., Zaitsev, I., Joachims, T.: A general framework for
  counterfactual learning-to-rank. In: Proceedings of the 42nd International
  ACM SIGIR Conference on Research and Development in Information Retrieval. p.
  5–14. SIGIR’19, Association for Computing Machinery, New York, NY, USA
  (2019). \doi{10.1145/3331184.3331202},
  \url{https://doi.org/10.1145/3331184.3331202}

\bibitem{Bilgic05}
Bilgic, M., Mooney, R.J.: Explaining recommendations: Satisfaction vs.
  promotion. In: Beyond Personalization Workshop, IUI. vol.~5, p.~153 (2005)

\bibitem{Bodapati08}
Bodapati, A.V.: Recommendation systems with purchase data. Journal of marketing
  research  \textbf{45}(1),  77--93 (2008)

\bibitem{Bonner18}
Bonner, S., Vasile, F.: Causal embeddings for recommendation. In: Proceedings
  of the 12th ACM Conference on Recommender Systems. p. 104–112. RecSys
  ’18, Association for Computing Machinery, New York, NY, USA (2018).
  \doi{10.1145/3240323.3240360}, \url{https://doi.org/10.1145/3240323.3240360}

\bibitem{Bottou13}
Bottou, L., Peters, J., Qui\~{n}onero Candela, J., Charles, D.X., Chickering,
  D.M., Portugaly, E., Ray, D., Simard, P., Snelson, E.: Counterfactual
  reasoning and learning systems: The example of computational advertising. J.
  Mach. Learn. Res.  \textbf{14}(1),  3207–3260 (Jan 2013)

\bibitem{Dacrema19}
Dacrema, M.F., Cremonesi, P., Jannach, D.: Are we really making much progress?
  a worrying analysis of recent neural recommendation approaches. In:
  Proceedings of the 13th ACM Conference on Recommender Systems. p. 101–109.
  RecSys '19, Association for Computing Machinery, New York, NY, USA (2019).
  \doi{10.1145/3298689.3347058}, \url{https://doi.org/10.1145/3298689.3347058}

\bibitem{Devriendt18}
Devriendt, F., Moldovan, D., Verbeke, W.: A literature survey and experimental
  evaluation of the state-of-the-art in uplift modeling: A stepping stone
  toward the development of prescriptive analytics. Big data  \textbf{6}(1),
  13--41 (2018)

\bibitem{Fouss07}
Fouss, F., Pirotte, A., Renders, J.M., Saerens, M.: Random-walk computation of
  similarities between nodes of a graph with application to collaborative
  recommendation. IEEE Transactions on Knowledge and Data Engineering
  \textbf{19}(3),  355--369 (2007)

\bibitem{Harper15}
Harper, F.M., Konstan, J.A.: The movielens datasets: History and context. ACM
  Trans. Interact. Intell. Syst.  \textbf{5}(4) (Dec 2015).
  \doi{10.1145/2827872}, \url{https://doi.org/10.1145/2827872}

\bibitem{Hernan20}
Hern{\'a}n, M., Robins, J.: Causal inference: What if. Boca Raton: Chapman \&
  Hill/CRC  (2020)

\bibitem{Holland86}
Holland, P.W.: Statistics and causal inference. Journal of the American
  statistical Association  \textbf{81}(396),  945--960 (1986)

\bibitem{Hudgens08}
Hudgens, M.G., Halloran, M.E.: Toward causal inference with interference.
  Journal of the American Statistical Association  \textbf{103}(482),  832--842
  (2008)

\bibitem{Imbens15}
Imbens, G.W., Rubin, D.B.: Causal Inference for Statistics, Social, and
  Biomedical Sciences: An Introduction. Cambridge University Press, New York,
  NY, USA (2015)

\bibitem{Jannach19}
Jannach, D., Jugovac, M.: Measuring the business value of recommender systems.
  ACM Trans. Manage. Inf. Syst.  \textbf{10}(4) (Dec 2019).
  \doi{10.1145/3370082}, \url{https://doi.org/10.1145/3370082}

\bibitem{Jaskowski12}
Jaskowski, M., Jaroszewicz, S.: Uplift modeling for clinical trial data. In:
  ICML Workshop on Clinical Data Analysis (2012)

\bibitem{Joachims16}
Joachims, T., Swaminathan, A.: Counterfactual evaluation and learning for
  search, recommendation and ad placement. p. 1199–1201. SIGIR '16,
  Association for Computing Machinery, New York, NY, USA (2016).
  \doi{10.1145/2911451.2914803}, \url{https://doi.org/10.1145/2911451.2914803}

\bibitem{Joachims17}
Joachims, T., Swaminathan, A., Schnabel, T.: Unbiased learning-to-rank with
  biased feedback. In: Proceedings of the Tenth ACM International Conference on
  Web Search and Data Mining. p. 781–789. WSDM ’17, Association for
  Computing Machinery, New York, NY, USA (2017). \doi{10.1145/3018661.3018699},
  \url{https://doi.org/10.1145/3018661.3018699}

\bibitem{Johansson16}
Johansson, F.D., Shalit, U., Sontag, D.: Learning representations for
  counterfactual inference. In: Proceedings of the 33rd International
  Conference on International Conference on Machine Learning - Volume 48. p.
  3020–3029. ICML'16, JMLR.org (2016)

\bibitem{Johnson14}
Johnson, C.C.: Logistic matrix factorization for implicit feedback data.
  Advances in Neural Information Processing Systems  \textbf{27} (2014)

\bibitem{Kabbur13}
Kabbur, S., Ning, X., Karypis, G.: Fism: Factored item similarity models for
  top-n recommender systems. In: Proceedings of the 19th ACM SIGKDD
  International Conference on Knowledge Discovery and Data Mining. p.
  659–667. KDD '13, Association for Computing Machinery, New York, NY, USA
  (2013). \doi{10.1145/2487575.2487589},
  \url{https://doi.org/10.1145/2487575.2487589}

\bibitem{Kane14}
Kane, K., Lo, V.S., Zheng, J.: Mining for the truly responsive customers and
  prospects using true-lift modeling: Comparison of new and existing methods.
  Journal of Marketing Analytics  \textbf{2}(4),  218--238 (2014)

\bibitem{Koren15}
Koren, Y., Bell, R.: Advances in Collaborative Filtering, pp. 77--118. Springer
  US, Boston, MA (2015)

\bibitem{Koren09}
Koren, Y., Bell, R., Volinsky, C.: Matrix factorization techniques for
  recommender systems. Computer (8),  30--37 (2009)

\bibitem{Liang16}
Liang, D., Charlin, L., McInerney, J., Blei, D.M.: Modeling user exposure in
  recommendation. In: Proceedings of the 25th International Conference on World
  Wide Web. pp. 951--961. WWW '16 (2016)

\bibitem{Ludewig19}
Ludewig, M., Mauro, N., Latifi, S., Jannach, D.: Performance comparison of
  neural and non-neural approaches to session-based recommendation. In:
  Proceedings of the 13th ACM Conference on Recommender Systems. p. 462–466.
  RecSys '19, Association for Computing Machinery, New York, NY, USA (2019).
  \doi{10.1145/3298689.3347041}, \url{https://doi.org/10.1145/3298689.3347041}

\bibitem{Lunceford04}
Lunceford, J.K., Davidian, M.: Stratification and weighting via the propensity
  score in estimation of causal treatment effects: a comparative study.
  Statistics in medicine  \textbf{23}(19),  2937--2960 (2004)

\bibitem{Luo08}
Luo, H., Niu, C., Shen, R., Ullrich, C.: A collaborative filtering framework
  based on both local user similarity and global user similarity. Machine
  Learning  \textbf{72}(3),  231--245 (2008)

\bibitem{Ning15}
Ning, X., Desrosiers, C., Karypis, G.: A Comprehensive Survey of
  Neighborhood-Based Recommendation Methods, pp. 37--76. Springer US, Boston,
  MA (2015)

\bibitem{Ning11}
Ning, X., Karypis, G.: Slim: Sparse linear methods for top-n recommender
  systems. In: 2011 IEEE 11th International Conference on Data Mining. pp.
  497--506. IEEE (2011)

\bibitem{Oosterhuis20}
Oosterhuis, H., Jagerman, R., de~Rijke, M.: Unbiased learning to rank:
  Counterfactual and online approaches. In: Companion Proceedings of the Web
  Conference 2020. p. 299–300. WWW '20, Association for Computing Machinery,
  New York, NY, USA (2020). \doi{10.1145/3366424.3383107},
  \url{https://doi.org/10.1145/3366424.3383107}

\bibitem{Radcliffe11}
Radcliffe, N.J., Surry, P.D.: Real-world uplift modelling with
  significance-based uplift trees. White Paper TR-2011-1, Stochastic Solutions
  (2011)

\bibitem{Rendle09}
Rendle, S., Freudenthaler, C., Gantner, Z., Schmidt-Thieme, L.: Bpr: Bayesian
  personalized ranking from implicit feedback. In: Proceedings of the
  Twenty-Fifth Conference on Uncertainty in Artificial Intelligence. p.
  452–461. UAI ’09, AUAI Press, Arlington, Virginia, USA (2009)

\bibitem{Rubin74}
Rubin, D.B.: Estimating causal effects of treatments in randomized and
  nonrandomized studies. Journal of educational Psychology  \textbf{66}(5),
  ~688 (1974)

\bibitem{Saito20b}
Saito, Y.: Doubly robust estimator for ranking metrics with post-click
  conversions. In: Fourteenth ACM Conference on Recommender Systems. p.
  92–100. RecSys '20, Association for Computing Machinery, New York, NY, USA
  (2020). \doi{10.1145/3383313.3412262},
  \url{https://doi.org/10.1145/3383313.3412262}

\bibitem{Saito20a}
Saito, Y., Yaginuma, S., Nishino, Y., Sakata, H., Nakata, K.: Unbiased
  recommender learning from missing-not-at-random implicit feedback. In:
  Proceedings of the 13th International Conference on Web Search and Data
  Mining. p. 501–509. WSDM ’20, Association for Computing Machinery, New
  York, NY, USA (2020). \doi{10.1145/3336191.3371783},
  \url{https://doi.org/10.1145/3336191.3371783}

\bibitem{Sarwar01}
Sarwar, B., Karypis, G., Konstan, J., Riedl, J.: Item-based collaborative
  filtering recommendation algorithms. In: Proceedings of the 10th
  International Conference on World Wide Web. p. 285–295. WWW '01,
  Association for Computing Machinery, New York, NY, USA (2001).
  \doi{10.1145/371920.372071}, \url{https://doi.org/10.1145/371920.372071}

\bibitem{Sato16}
Sato, M., Izumo, H., Sonoda, T.: Modeling individual users' responsiveness to
  maximize recommendation impact. In: Proceedings of the 2016 Conference on
  User Modeling Adaptation and Personalization. pp. 259--267. UMAP '16, ACM,
  New York, NY, USA (2016). \doi{10.1145/2930238.2930259},
  \url{http://doi.acm.org/10.1145/2930238.2930259}

\bibitem{Sato19}
Sato, M., Singh, J., Takemori, S., Sonoda, T., Zhang, Q., Ohkuma, T.:
  Uplift-based evaluation and optimization of recommenders. In: Proceedings of
  the 13th ACM Conference on Recommender Systems. p. 296–304. RecSys ’19,
  Association for Computing Machinery, New York, NY, USA (2019).
  \doi{10.1145/3298689.3347018}, \url{https://doi.org/10.1145/3298689.3347018}

\bibitem{Sato20}
Sato, M., Takemori, S., Singh, J., Ohkuma, T.: Unbiased learning for the causal
  effect of recommendation. In: Fourteenth ACM Conference on Recommender
  Systems. p. 378–387. RecSys '20, Association for Computing Machinery, New
  York, NY, USA (2020). \doi{10.1145/3383313.3412261},
  \url{https://doi.org/10.1145/3383313.3412261}

\bibitem{Schnabel16}
Schnabel, T., Swaminathan, A., Singh, A., Chandak, N., Joachims, T.:
  Recommendations as treatments: Debiasing learning and evaluation. In:
  Proceedings of the 33rd International Conference on International Conference
  on Machine Learning - Volume 48. p. 1670–1679. ICML’16, JMLR.org (2016)

\bibitem{Shardanand95}
Shardanand, U., Maes, P.: Social information filtering: Algorithms for
  automating “word of mouth”. In: Proceedings of the SIGCHI Conference on
  Human Factors in Computing Systems. p. 210–217. CHI '95, ACM
  Press/Addison-Wesley Publishing Co., USA (1995). \doi{10.1145/223904.223931},
  \url{https://doi.org/10.1145/223904.223931}

\bibitem{Sharma15}
Sharma, A., Hofman, J.M., Watts, D.J.: Estimating the causal impact of
  recommendation systems from observational data. In: Proceedings of the
  Sixteenth ACM Conference on Economics and Computation. pp. 453--470. EC '15,
  ACM, New York, NY, USA (2015). \doi{10.1145/2764468.2764488},
  \url{http://doi.acm.org/10.1145/2764468.2764488}

\bibitem{Stuart10}
Stuart, E.A.: Matching methods for causal inference: A review and a look
  forward. Statist. Sci.  \textbf{25}(1),  1--21 (02 2010).
  \doi{10.1214/09-STS313}, \url{https://doi.org/10.1214/09-STS313}

\bibitem{Swaminathan15b}
Swaminathan, A., Joachims, T.: The self-normalized estimator for counterfactual
  learning. In: Advances in Neural Information Processing Systems 28: Annual
  Conference on Neural Information Processing Systems 2015, December 7-12,
  2015, Montreal, Quebec, Canada. pp. 3231--3239 (2015),
  \url{http://papers.nips.cc/paper/5748-the-self-normalized-estimator-for-counterfactual-learning}

\bibitem{Tintarev08}
Tintarev, N., Masthoff, J.: Over- and underestimation in different product
  domains. In: Ghallab, M., Spyropoulos, C., Fakotakis, N., Avouris, N. (eds.)
  Workshop on Recommender Systems. IOS Press (Jul 2008), workshop on
  Recommender Systems, 18th European Conference on Artificial Intelligence ;
  18th European Conference on Artificial Intelligence (ECAI 2008) ; Conference
  date: 21-07-2008 Through 25-07-2008

\bibitem{Tintarev2015}
Tintarev, N., Masthoff, J.: Explaining Recommendations: Design and Evaluation,
  pp. 353--382. Springer US, Boston, MA (2015)

\bibitem{Torkamaan19}
Torkamaan, H., Barbu, C.M., Ziegler, J.: How can they know that? a study of
  factors affecting the creepiness of recommendations. In: Proceedings of the
  13th ACM Conference on Recommender Systems. p. 423–427. RecSys '19,
  Association for Computing Machinery, New York, NY, USA (2019).
  \doi{10.1145/3298689.3346982}, \url{https://doi.org/10.1145/3298689.3346982}

\bibitem{Tyler13}
Tyler, J.V., Miguel, A.H., et~al.: Causal inference under multiple versions of
  treatment. Journal of Causal Inference  \textbf{1}(1),  1--20 (2013)

\bibitem{Wang19}
Wang, X., Zhang, R., Sun, Y., Qi, J.: Doubly robust joint learning for
  recommendation on data missing not at random. In: Chaudhuri, K.,
  Salakhutdinov, R. (eds.) Proceedings of the 36th International Conference on
  Machine Learning. Proceedings of Machine Learning Research, vol.~97, pp.
  6638--6647. PMLR, Long Beach, California, USA (09--15 Jun 2019),
  \url{http://proceedings.mlr.press/v97/wang19n.html}

\bibitem{Wang16}
Wang, X., Bendersky, M., Metzler, D., Najork, M.: Learning to rank with
  selection bias in personal search. In: Proceedings of the 39th International
  ACM SIGIR Conference on Research and Development in Information Retrieval. p.
  115–124. SIGIR ’16, Association for Computing Machinery, New York, NY,
  USA (2016). \doi{10.1145/2911451.2911537},
  \url{https://doi.org/10.1145/2911451.2911537}

\bibitem{Wang19b}
Wang, Y., Blei, D.M.: The blessings of multiple causes. Journal of the American
  Statistical Association  \textbf{114}(528),  1574--1596 (2019)

\bibitem{Zhang19}
Zhang, S., Yao, L., Sun, A., Tay, Y.: Deep learning based recommender system: A
  survey and new perspectives. ACM Comput. Surv.  \textbf{52}(1) (Feb 2019).
  \doi{10.1145/3285029}, \url{https://doi.org/10.1145/3285029}

\bibitem{Zhuang20}
Zhuang, S., Zuccon, G.: Counterfactual online learning to rank. In: Jose, J.M.,
  Yilmaz, E., Magalh{\~a}es, J., Castells, P., Ferro, N., Silva, M.J., Martins,
  F. (eds.) Advances in Information Retrieval. pp. 415--430. Springer
  International Publishing, Cham (2020)

\end{thebibliography}

\appendix
\newpage
\section{Appendix}

\subsection{Additional Experimental Results}

\begin{figure*}[htbp]
	\begin{center}
		\subfigure[DH-Ori.]{\includegraphics[width=0.24\textwidth]{figures/fig_num_neighbor_dh_cate_original_CP10.png}}
		\subfigure[DH-Per.]{\includegraphics[width=0.24\textwidth]{figures/fig_num_neighbor_dh_cate_personalized_CP10v2.png}}
		\subfigure[ML-100K.]{\includegraphics[width=0.24\textwidth]{figures/fig_num_neighbor_ml100k_CP10.png}}
		\subfigure[ML-1M.]{\includegraphics[width=0.24\textwidth]{figures/fig_num_neighbor_ml1m_CP10.png}}
		
		\subfigure[DH-Ori.]{\includegraphics[width=0.24\textwidth]{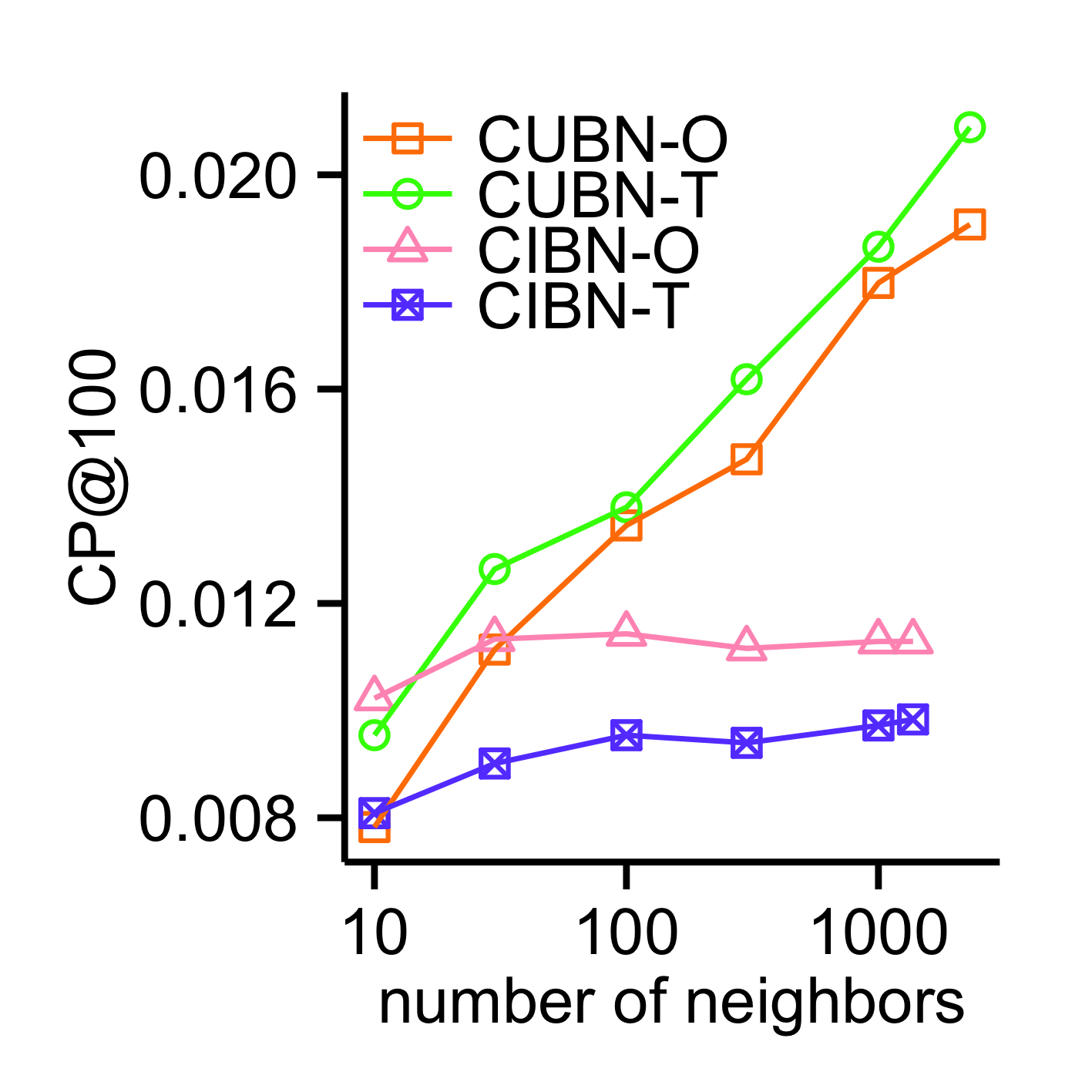}}
		\subfigure[DH-Per.]{\includegraphics[width=0.24\textwidth]{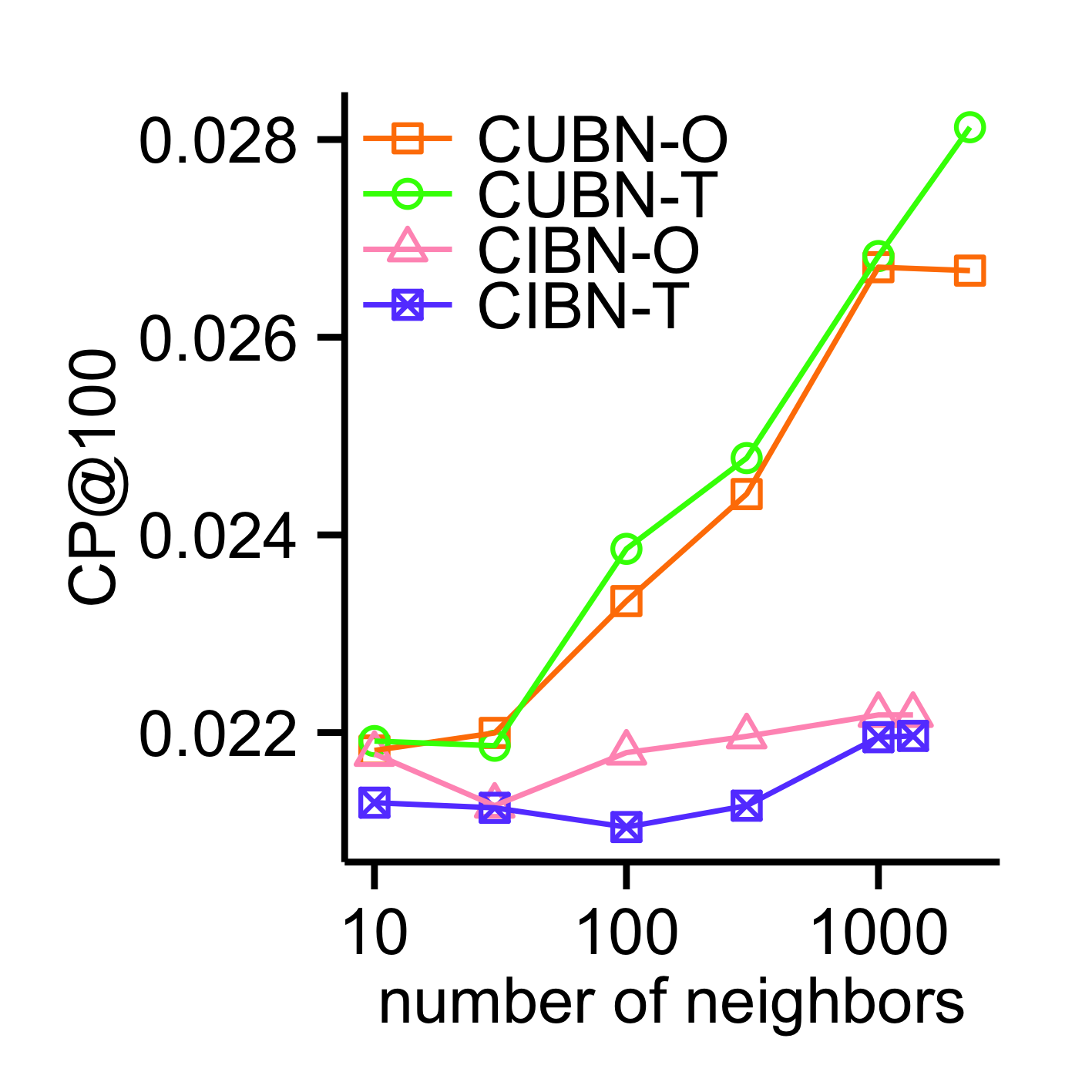}}
		\subfigure[ML-100K.]{\includegraphics[width=0.24\textwidth]{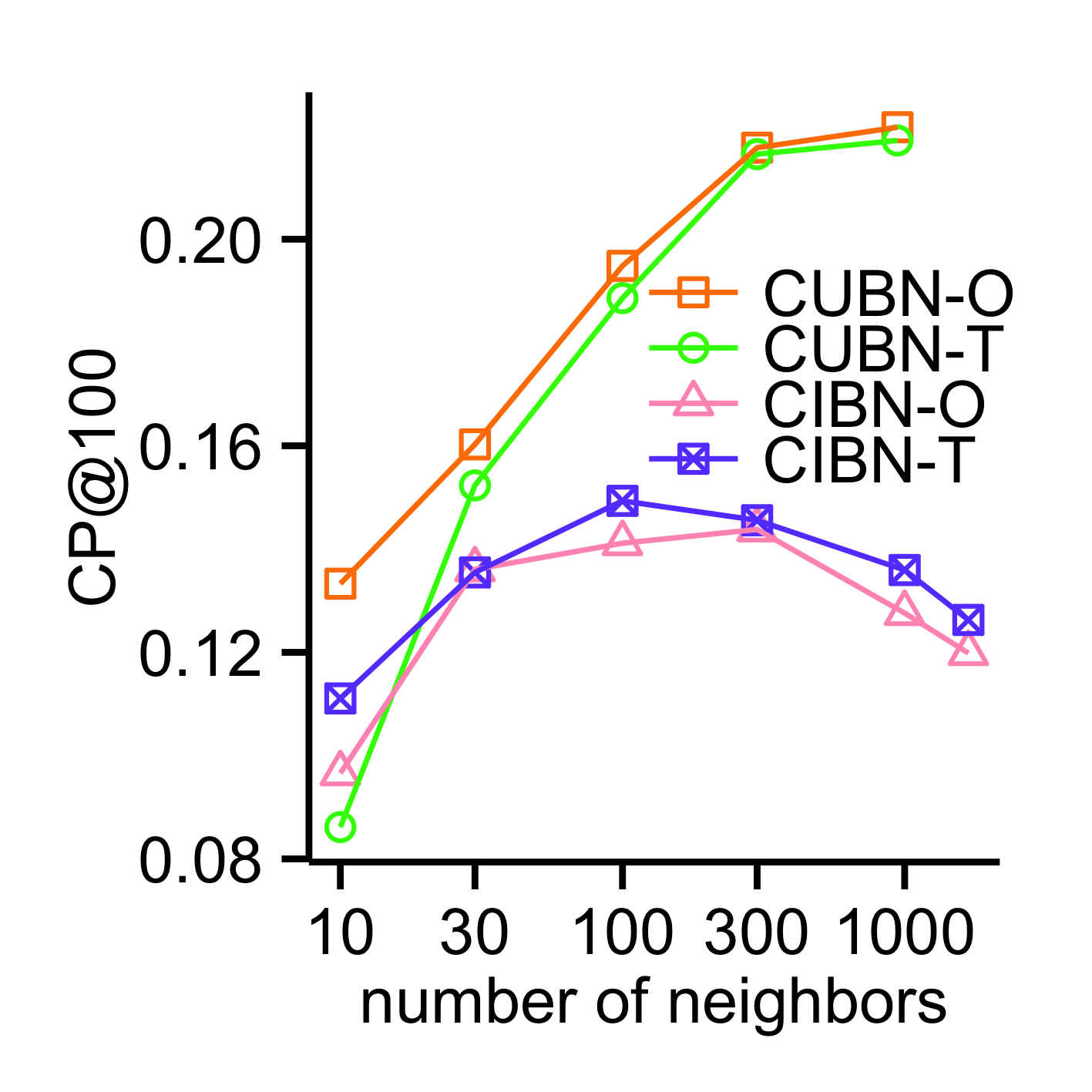}}
		\subfigure[ML-1M.]{\includegraphics[width=0.24\textwidth]{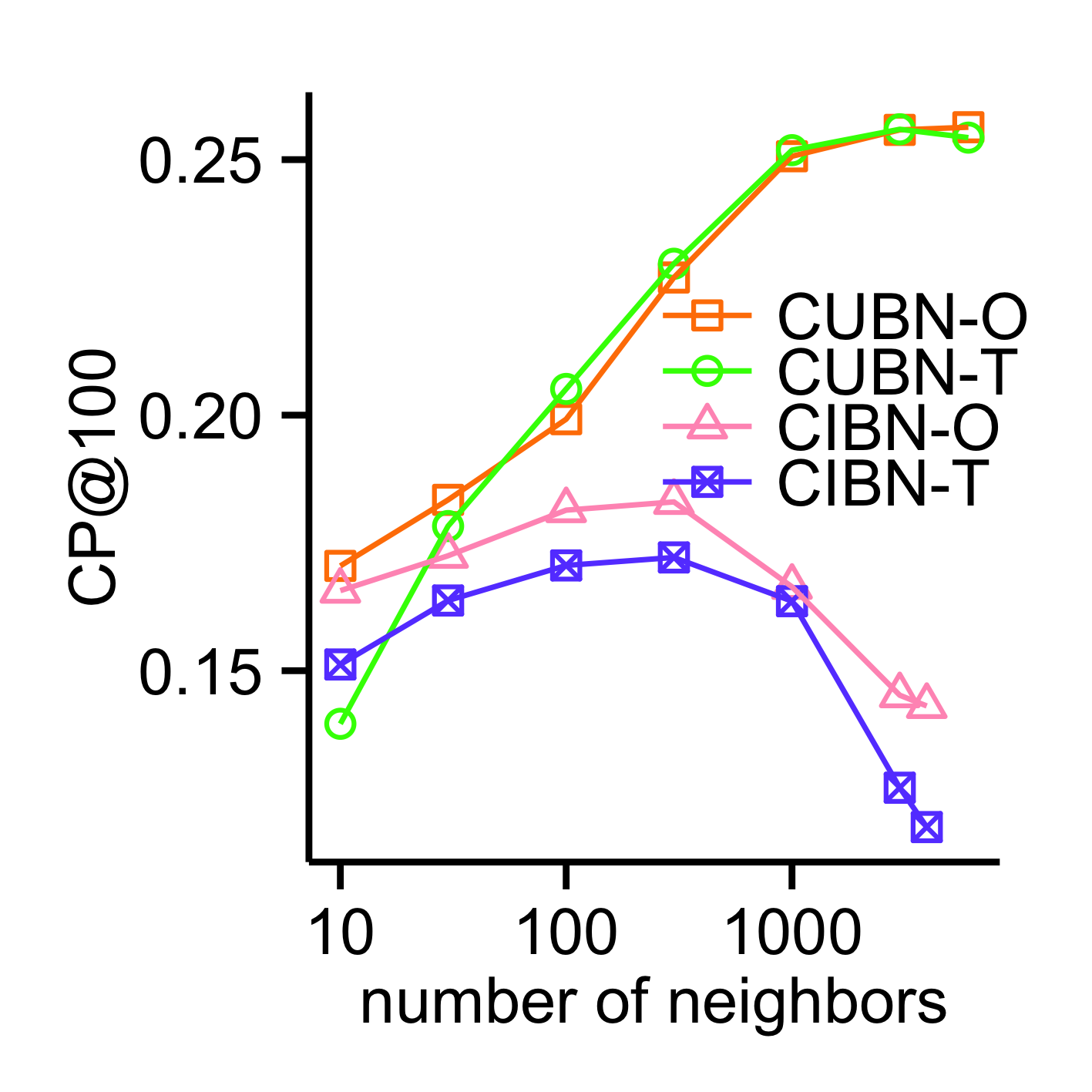}}
		
		\subfigure[DH-Ori.]{\includegraphics[width=0.24\textwidth]{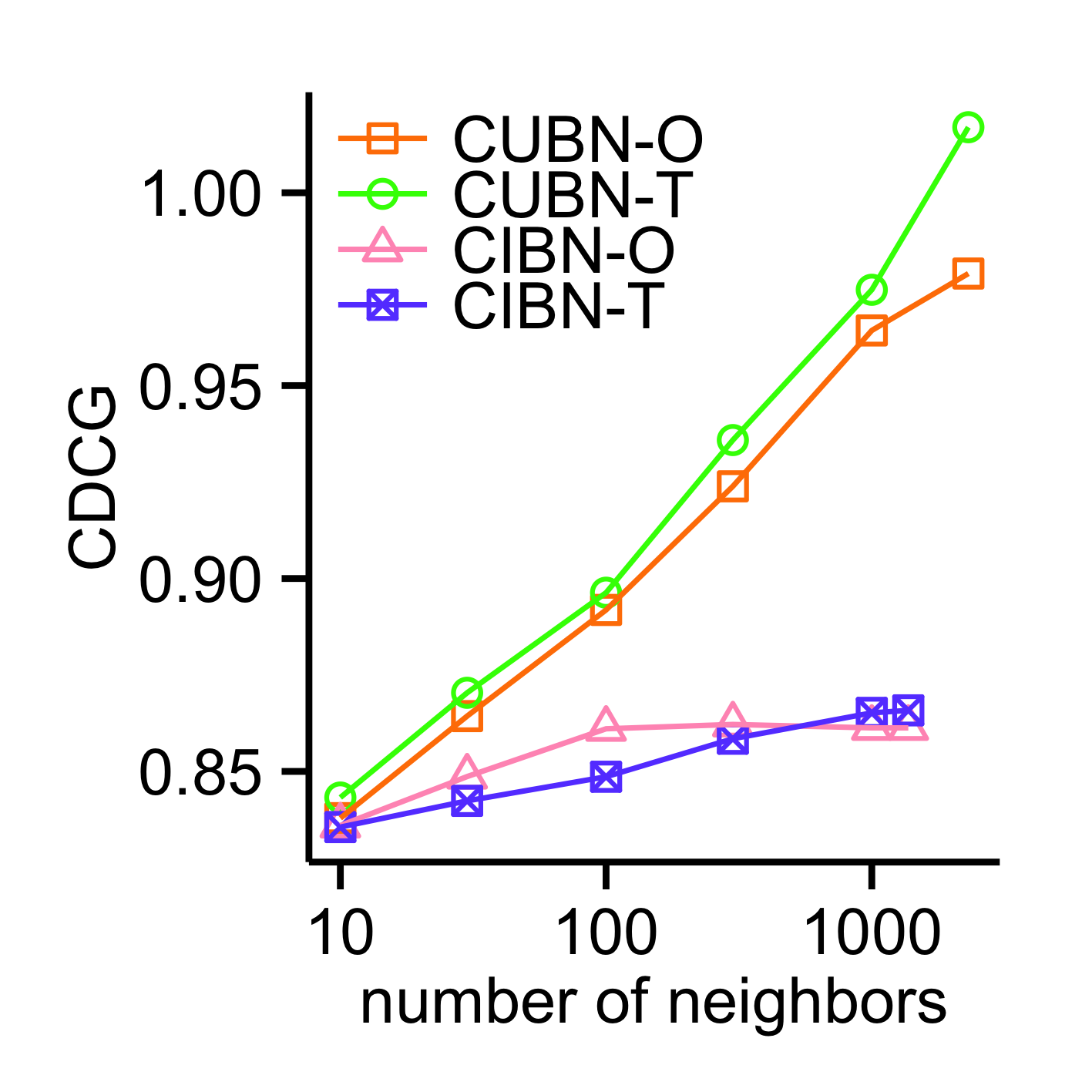}}
		\subfigure[DH-Per.]{\includegraphics[width=0.24\textwidth]{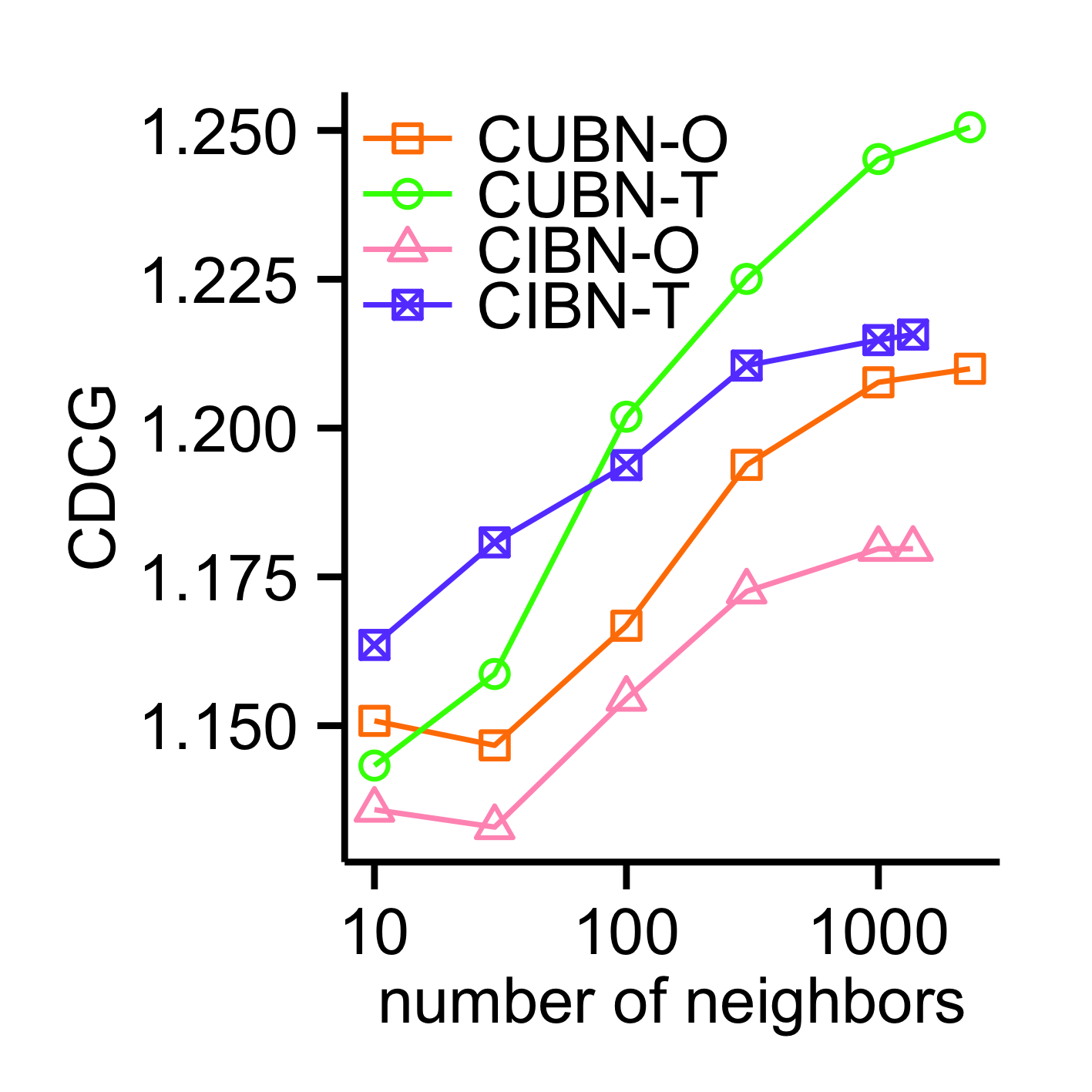}}
		\subfigure[ML-100K.]{\includegraphics[width=0.24\textwidth]{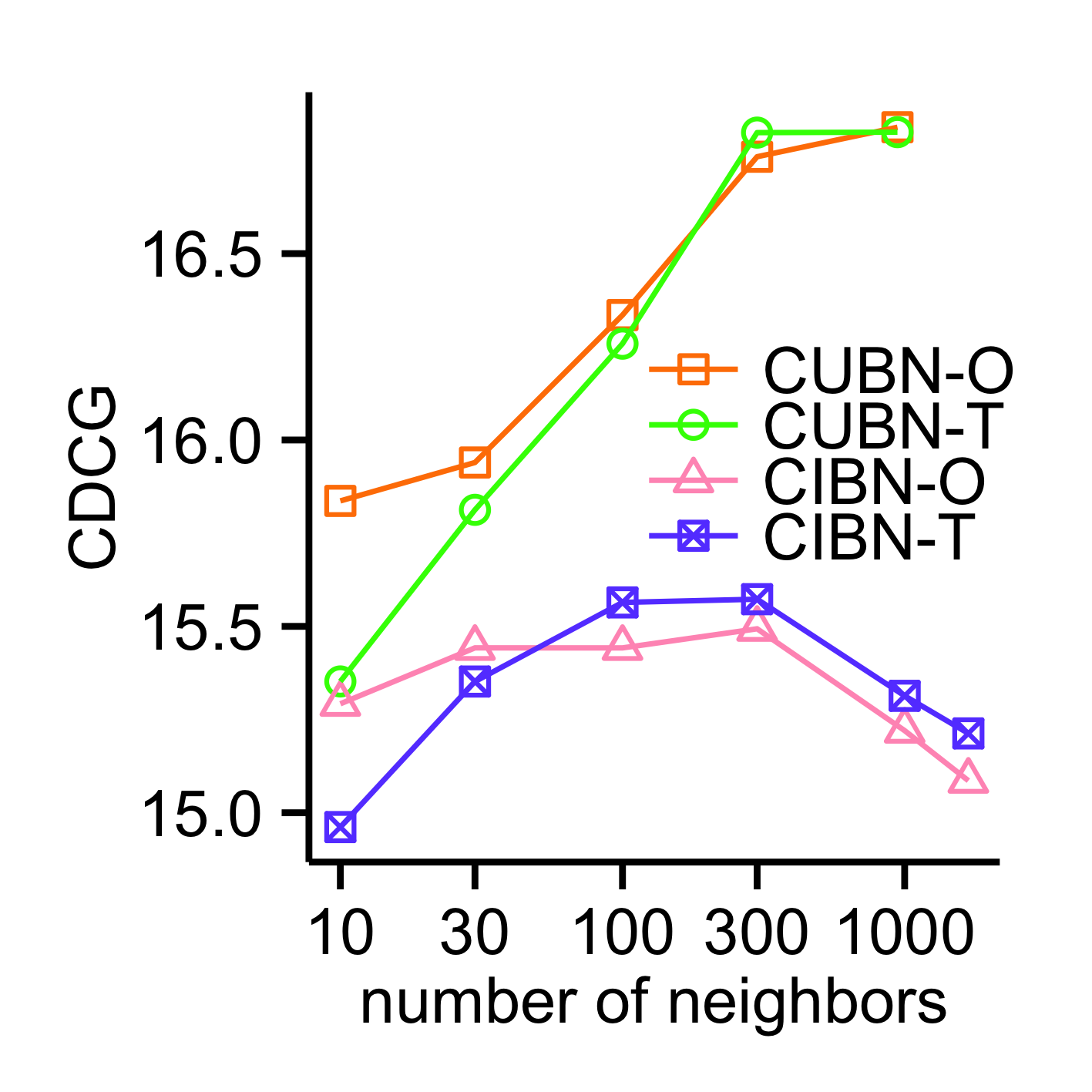}}
		\subfigure[ML-1M.]{\includegraphics[width=0.24\textwidth]{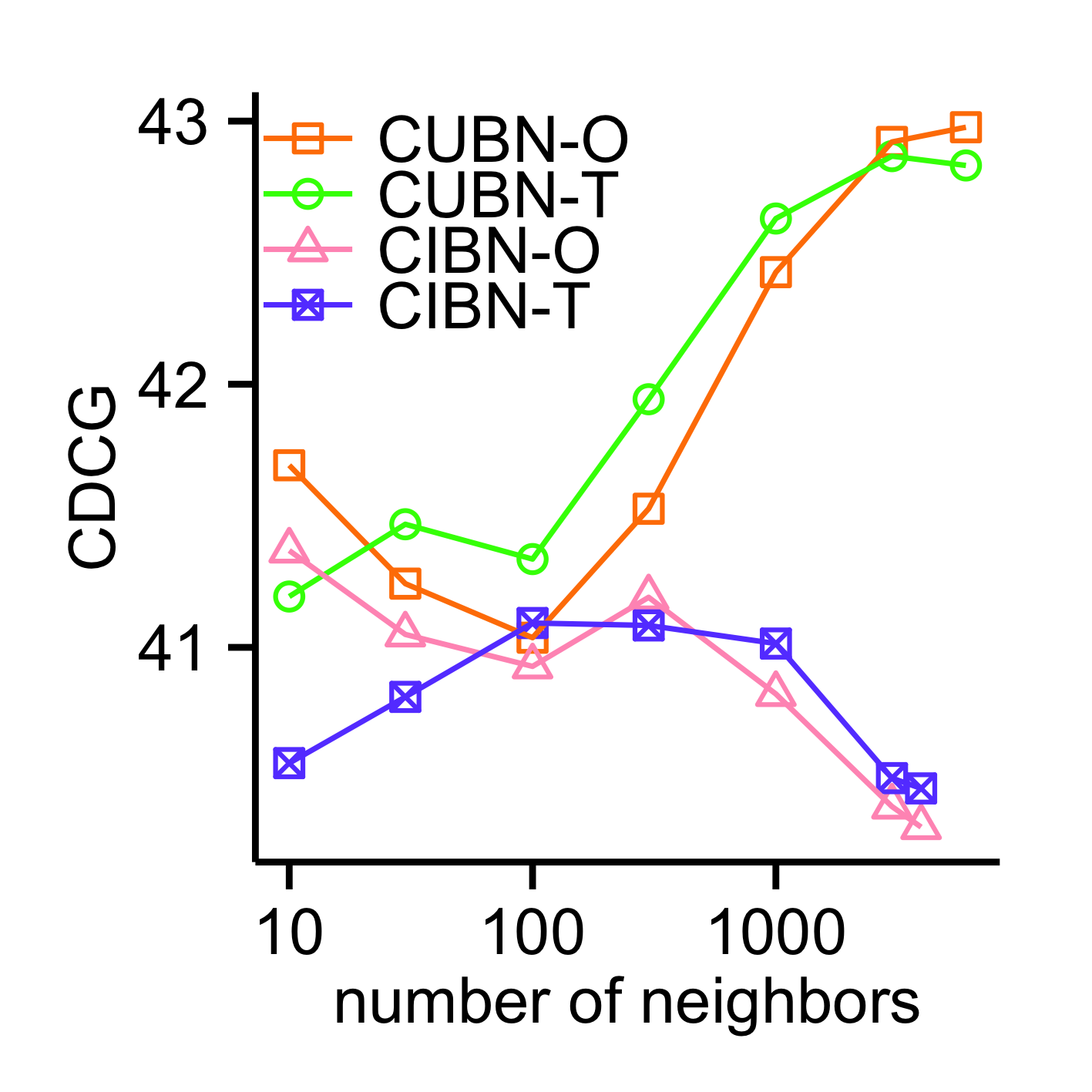}}
		\caption{Dependence on the number of neighbors in validation datasets. 
			The first, second, and third rows are results for CP@10, CP@100, and CDCG, respectively.
			The scaling factor $\alpha$ and the shrinkage parameter $\beta$ are set to the optimal values for each number of neighbors. Note that the possible number of neighbors are restricted by either the number of users or that of items.}
		\label{fig:num_neighbor_all}
	\end{center}
\end{figure*}

\begin{figure*}[htbp]
	\begin{center}
		\subfigure[CUBN-O (CP@10).]{\includegraphics[width=0.35\textwidth]{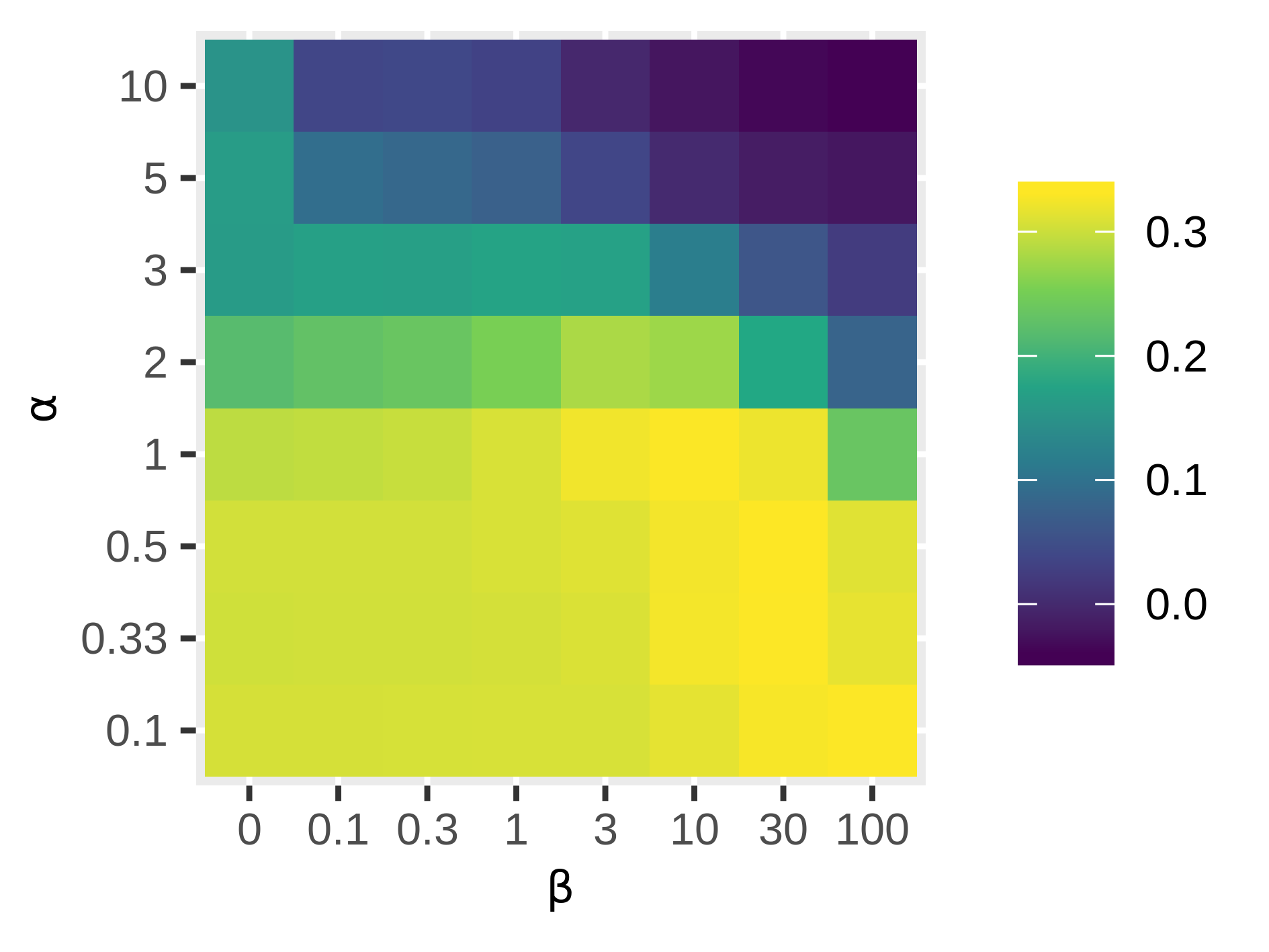}}
		\subfigure[CUBN-O (CP@100).]{\includegraphics[width=0.35\textwidth]{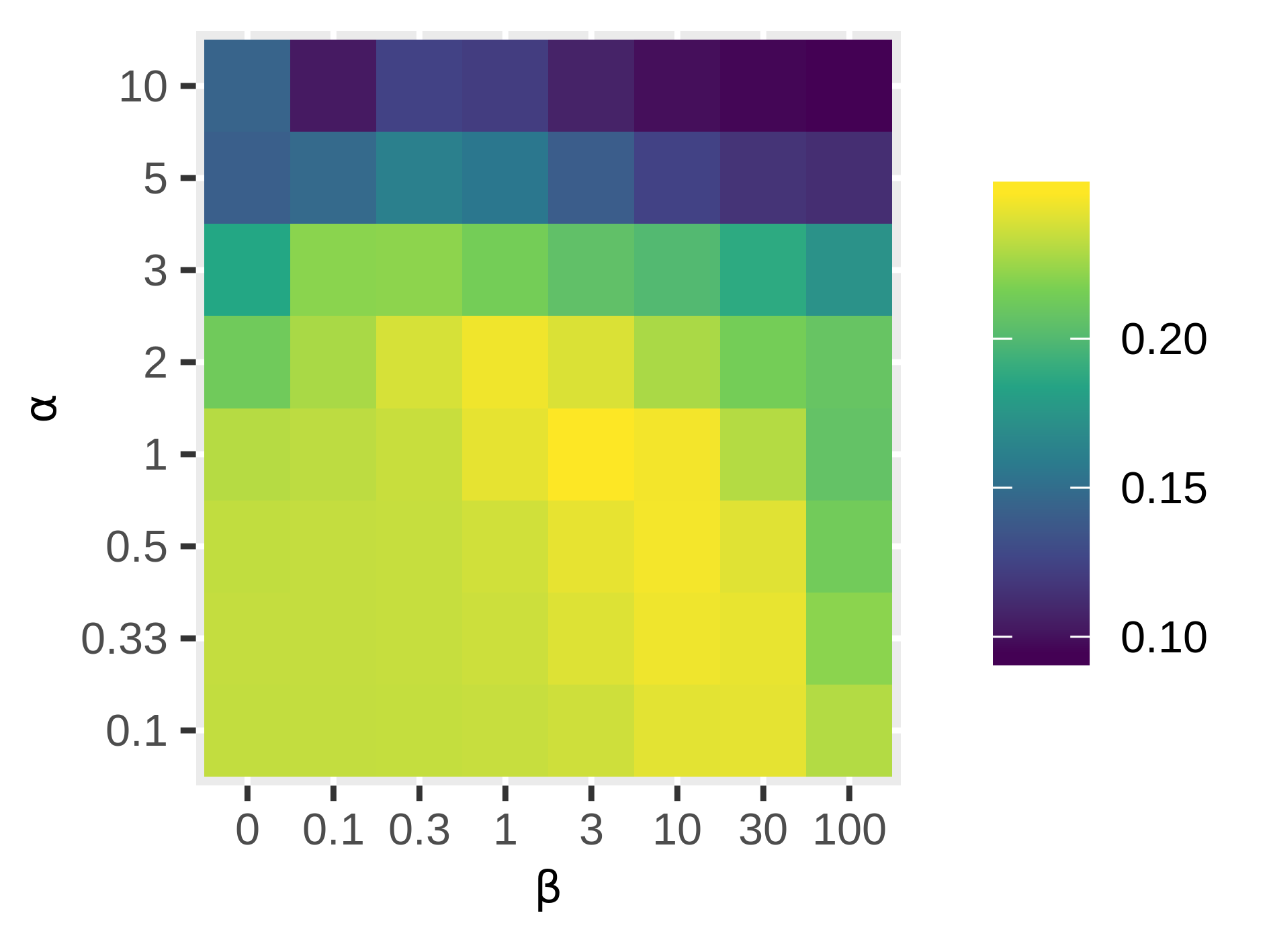}}
		\subfigure[CUBN-T (CP@10).]{\includegraphics[width=0.35\textwidth]{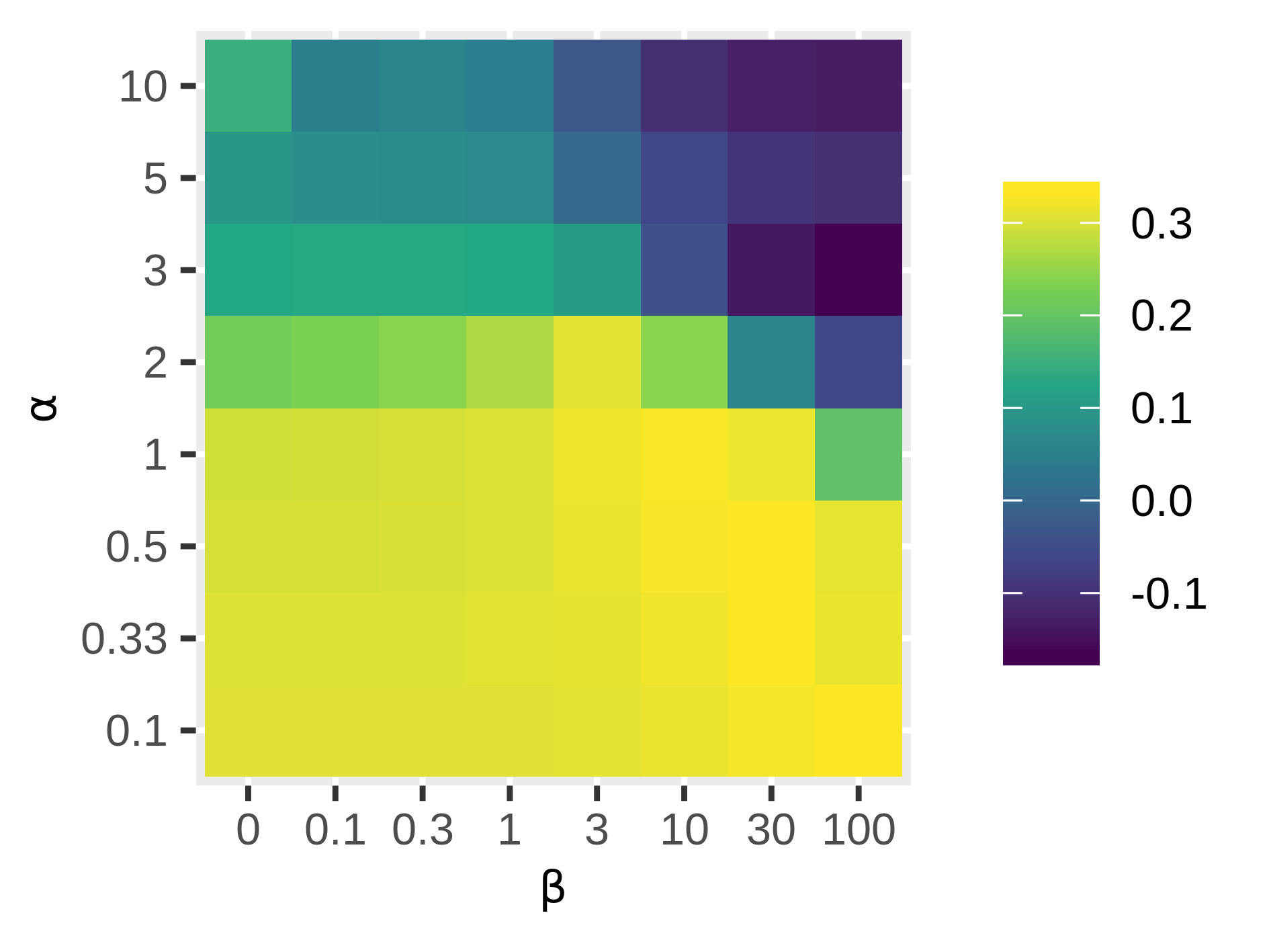}}
		\subfigure[CUBN-T (CP@100).]{\includegraphics[width=0.35\textwidth]{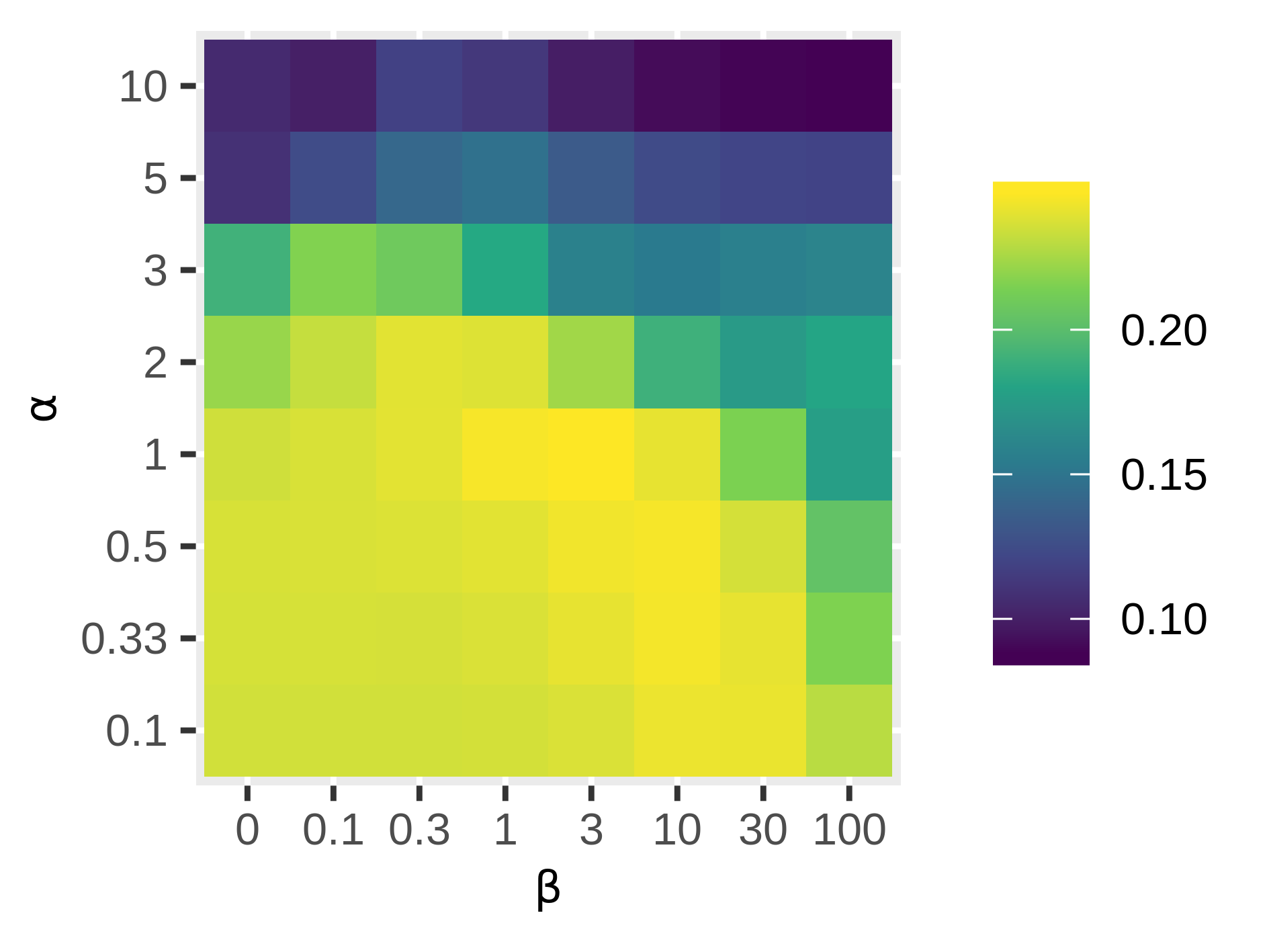}}
		\caption{Dependence on the scaling factor $\alpha$ and the shrinkage parameter $\beta$ in ML-1M.
			The number of neighbors are set to 6,040.}
		\label{fig:alpha_beta_ml1m}
	\end{center}
\end{figure*}

\begin{figure*}[htbp]
	\begin{center}
		\subfigure[CUBN-O (CP@10).]{\includegraphics[width=0.35\textwidth]{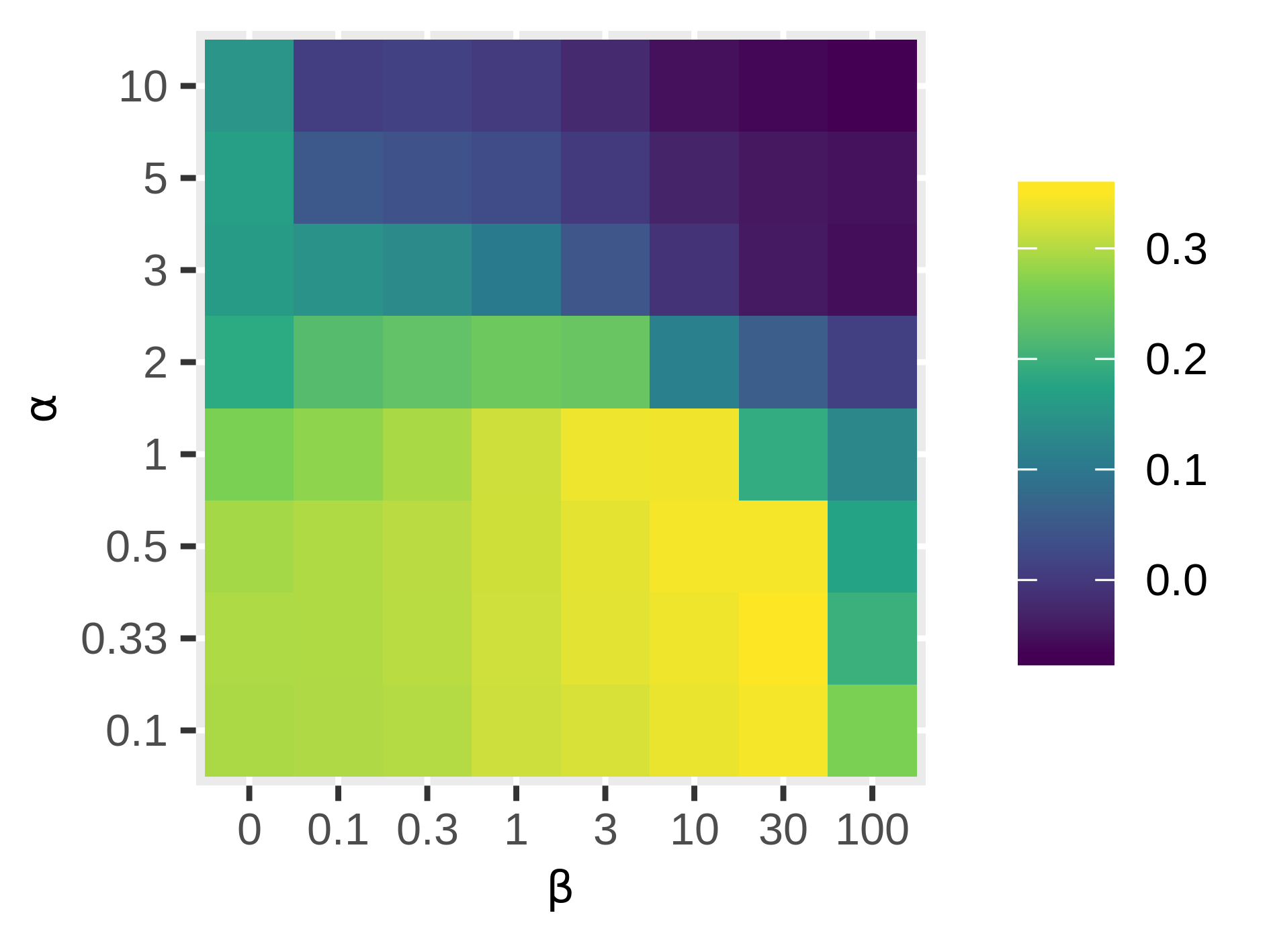}}
		\subfigure[CUBN-O (CP@100).]{\includegraphics[width=0.35\textwidth]{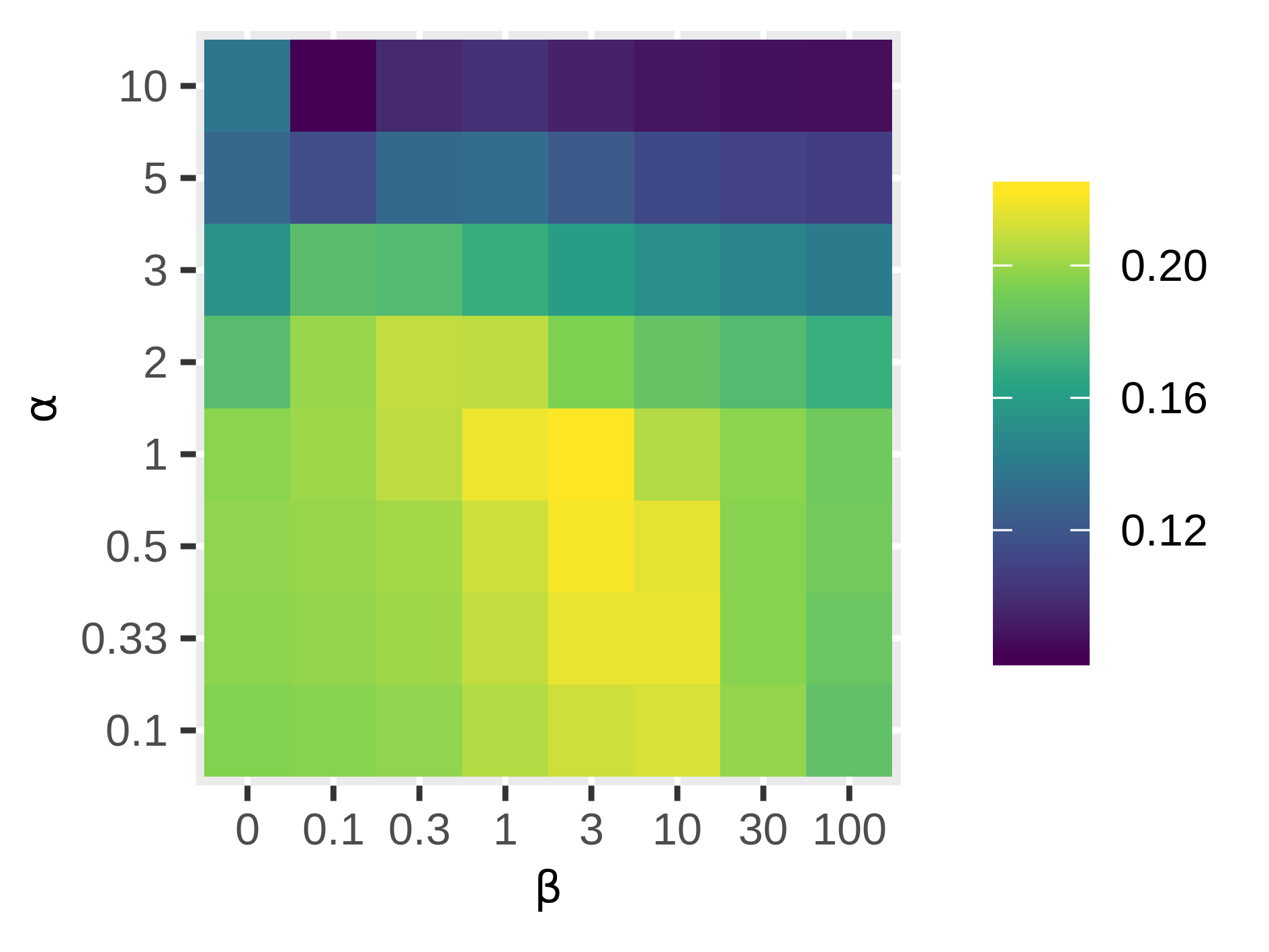}}
		\subfigure[CUBN-T (CP@10).]{\includegraphics[width=0.35\textwidth]{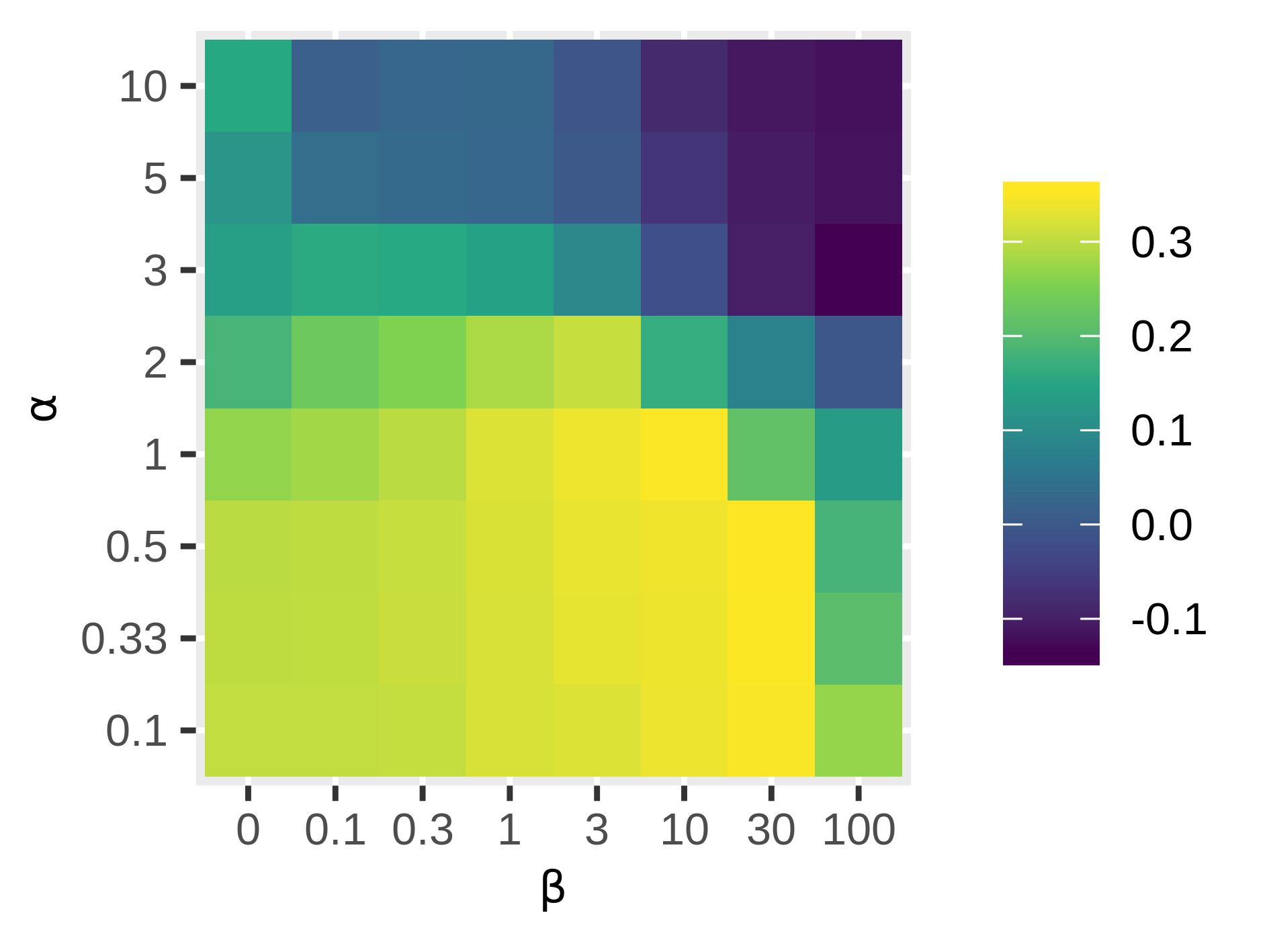}}
		\subfigure[CUBN-T (CP@100).]{\includegraphics[width=0.35\textwidth]{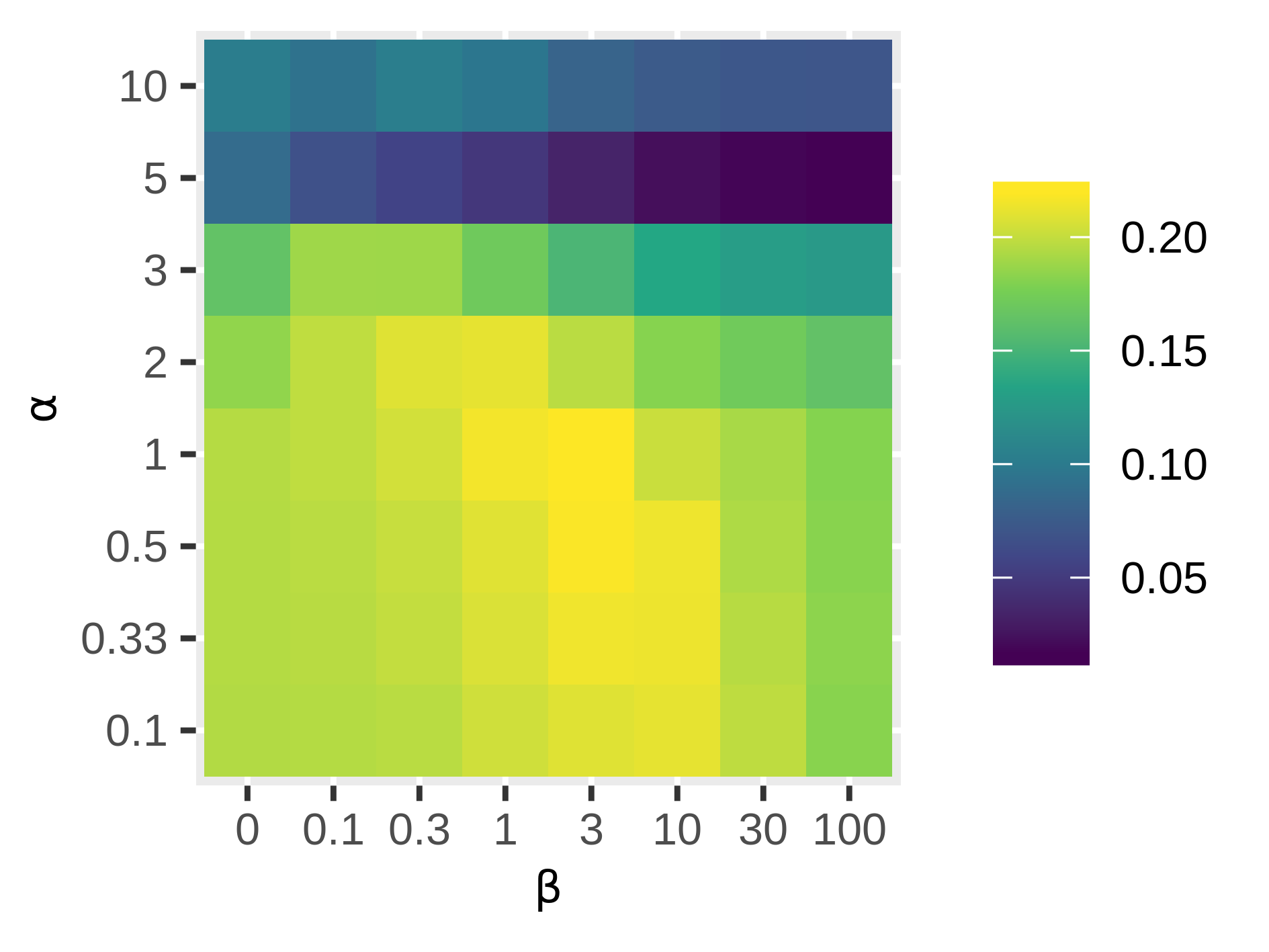}}
		\caption{Dependence on the scaling factor $\alpha$ and the shrinkage parameter $\beta$ in ML-100K.
			The number of neighbors are set to 943.}
		\label{fig:alpha_beta_ml100k}
	\end{center}
\end{figure*}

\begin{figure*}[htbp]
	\begin{center}
		\subfigure[CUBN-O (CP@10).]{\includegraphics[width=0.35\textwidth]{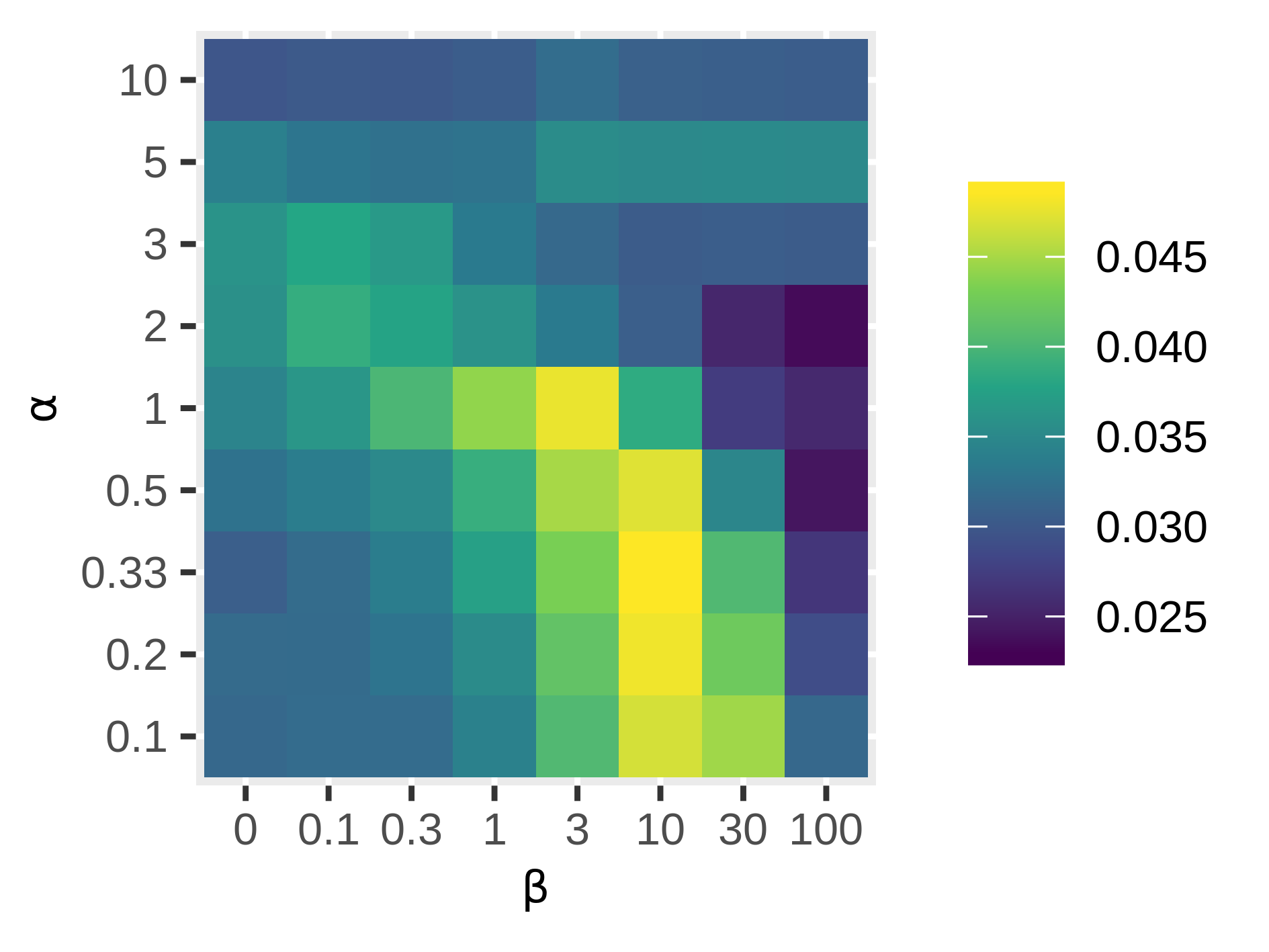}}
		\subfigure[CUBN-O (CP@100).]{\includegraphics[width=0.35\textwidth]{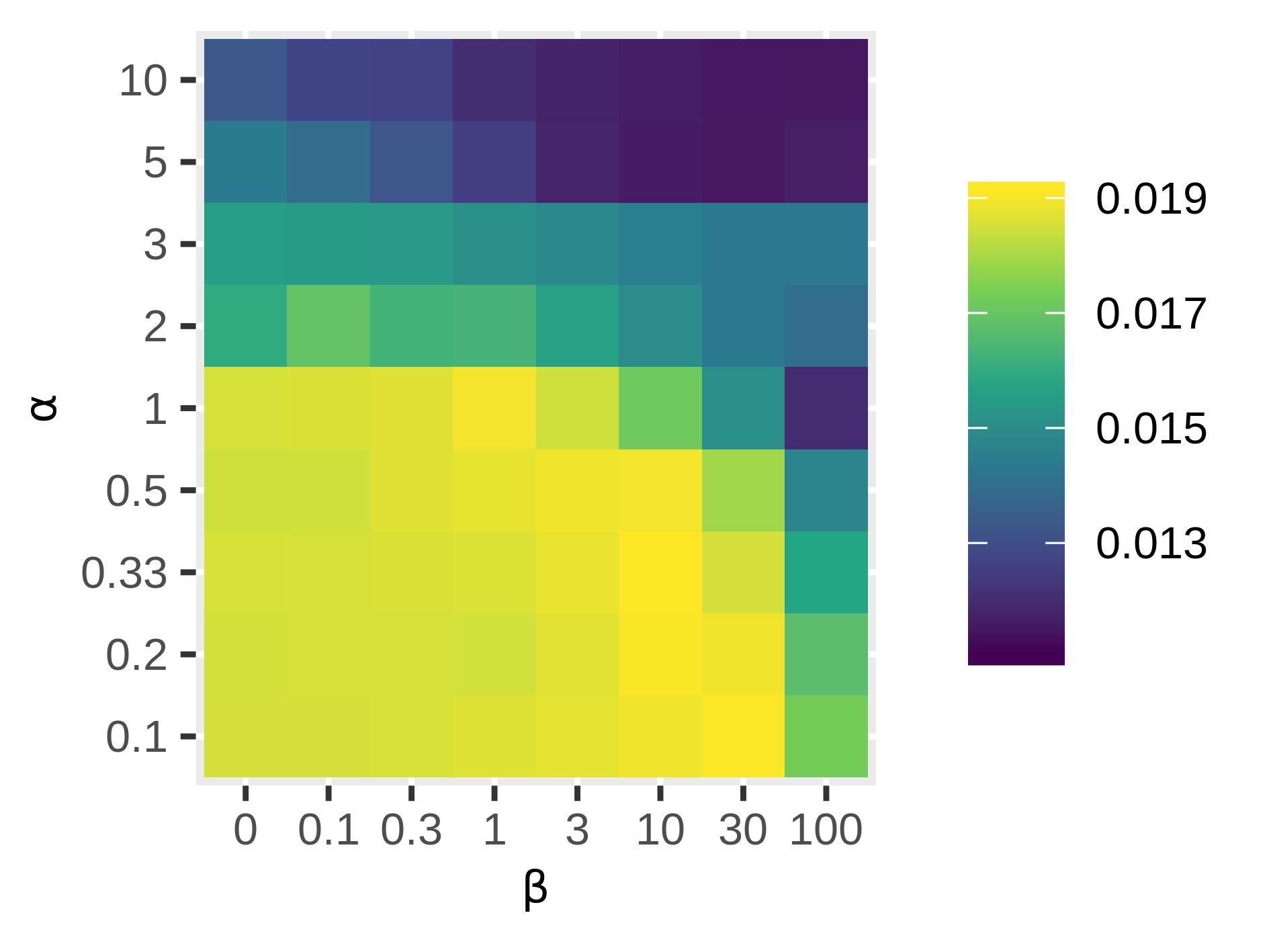}}
		\subfigure[CUBN-T (CP@10).]{\includegraphics[width=0.35\textwidth]{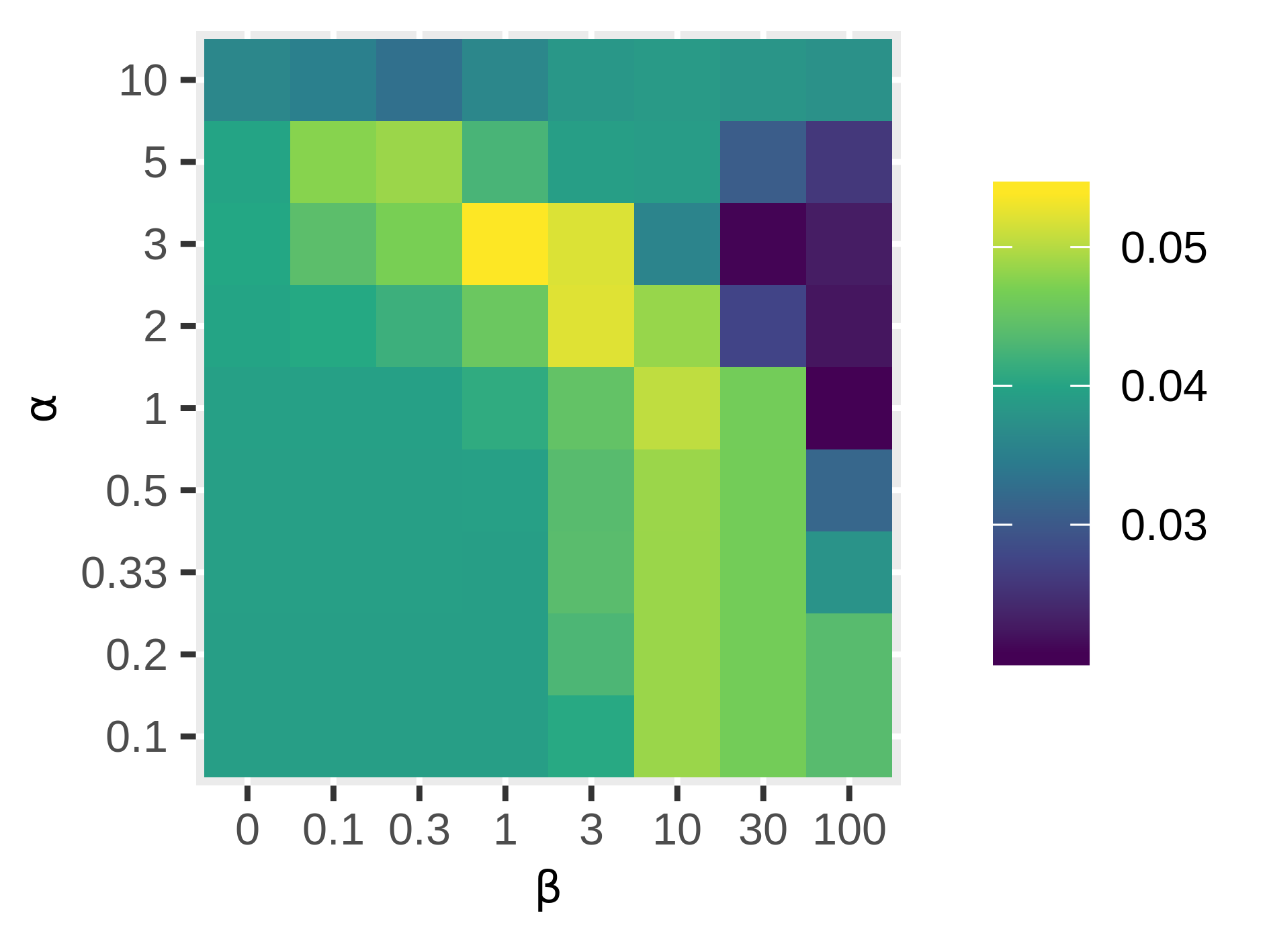}}
		\subfigure[CUBN-T (CP@100).]{\includegraphics[width=0.35\textwidth]{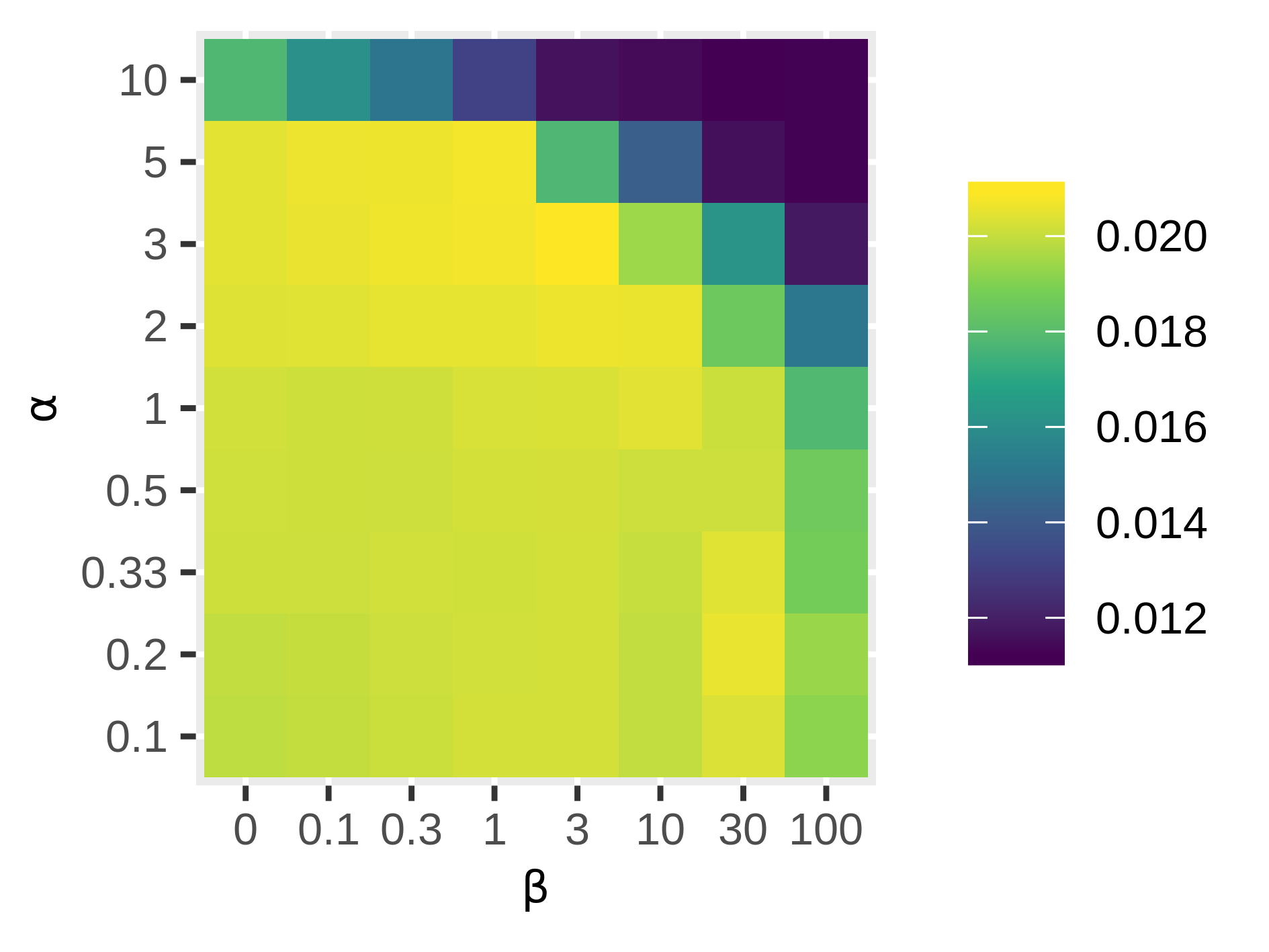}}
		\caption{Dependence on the scaling factor $\alpha$ and the shrinkage parameter $\beta$ in DH-Original.
			The number of neighbors are set to 2,309.}
		\label{fig:alpha_beta_dh_ori}
	\end{center}
\end{figure*}

\begin{figure*}[htbp]
	\begin{center}
		\subfigure[CUBN-O (CP@10).]{\includegraphics[width=0.35\textwidth]{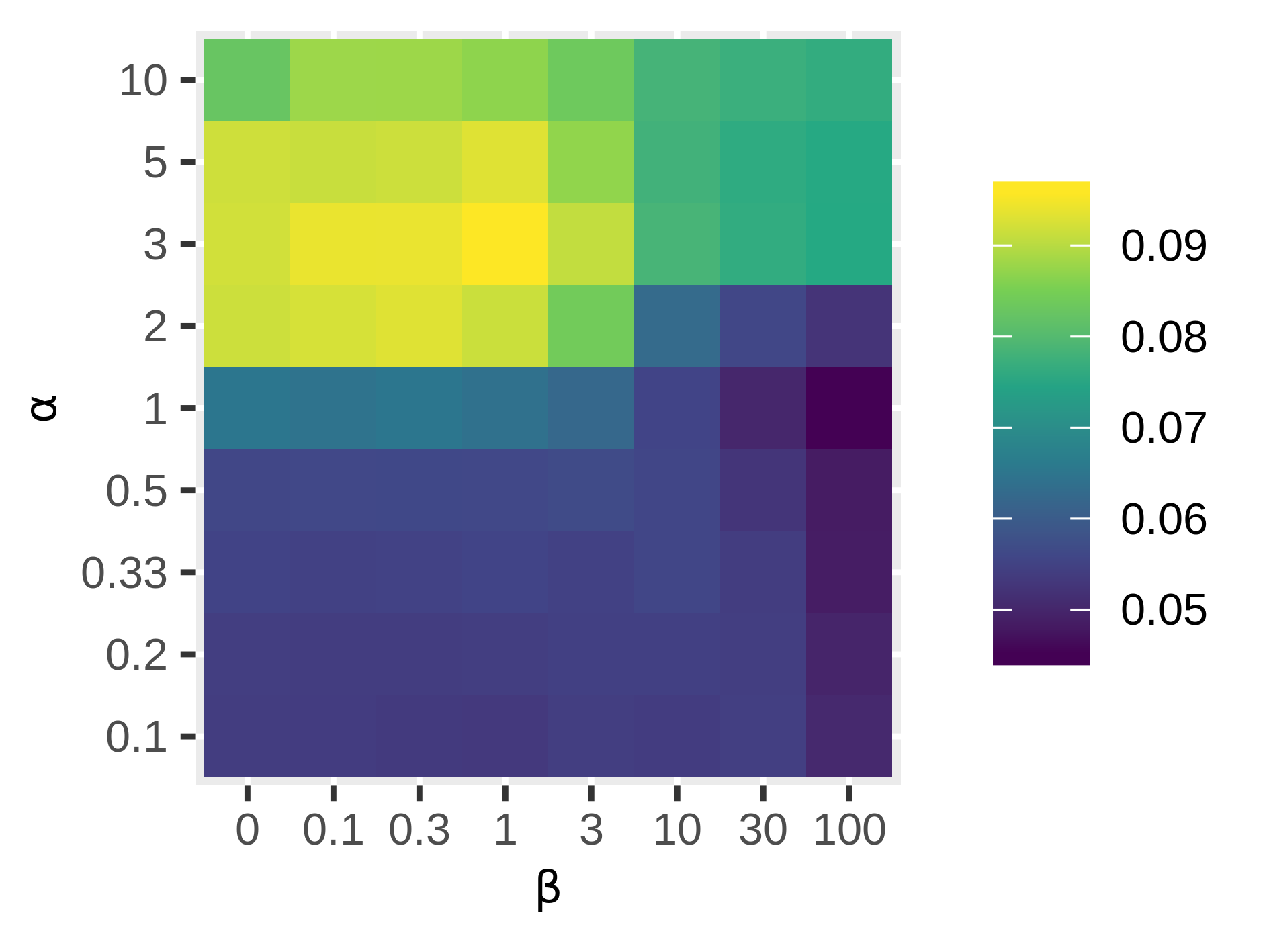}}
		\subfigure[CUBN-O (CP@100).]{\includegraphics[width=0.35\textwidth]{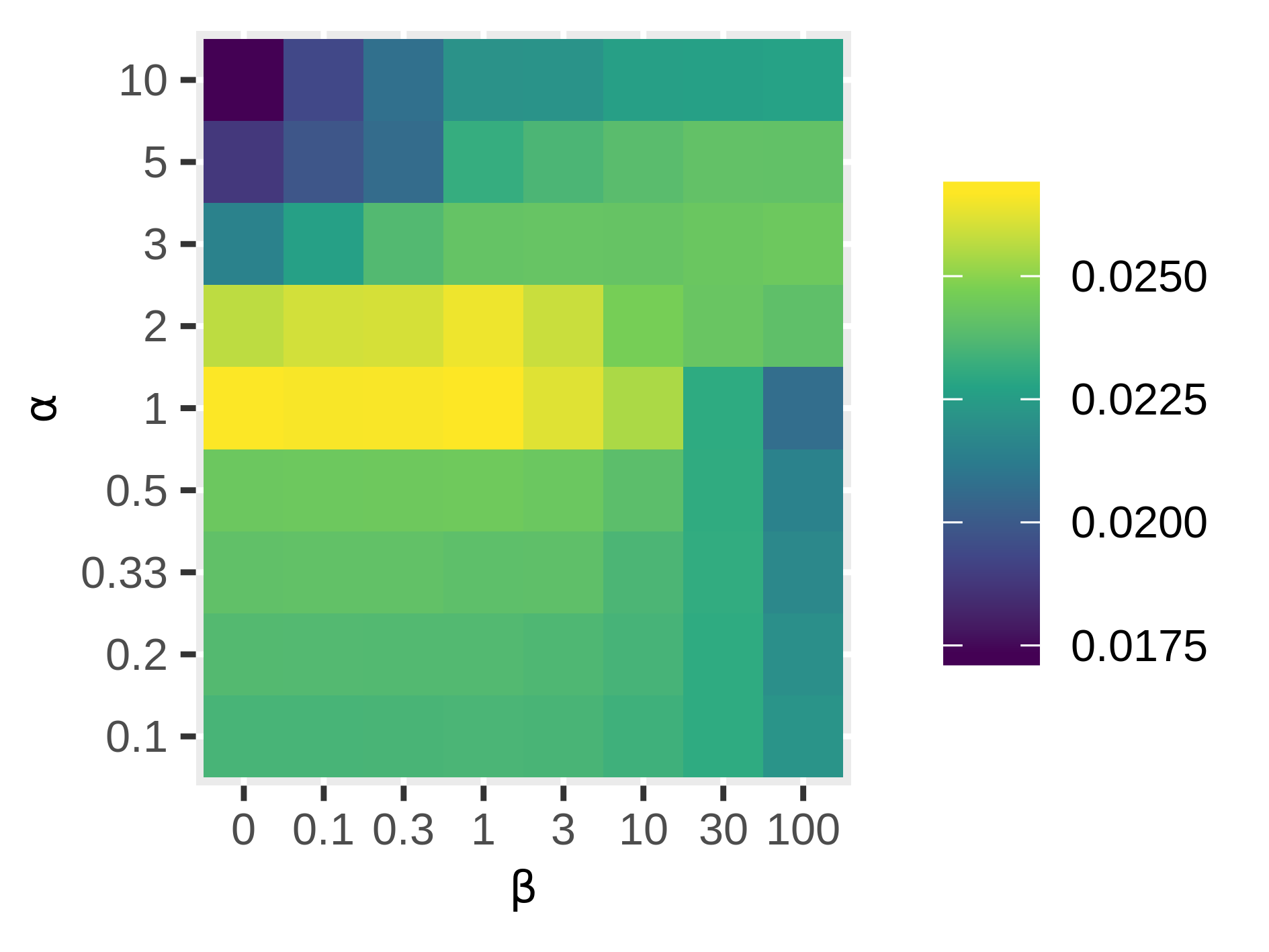}}
		\subfigure[CUBN-T (CP@10).]{\includegraphics[width=0.35\textwidth]{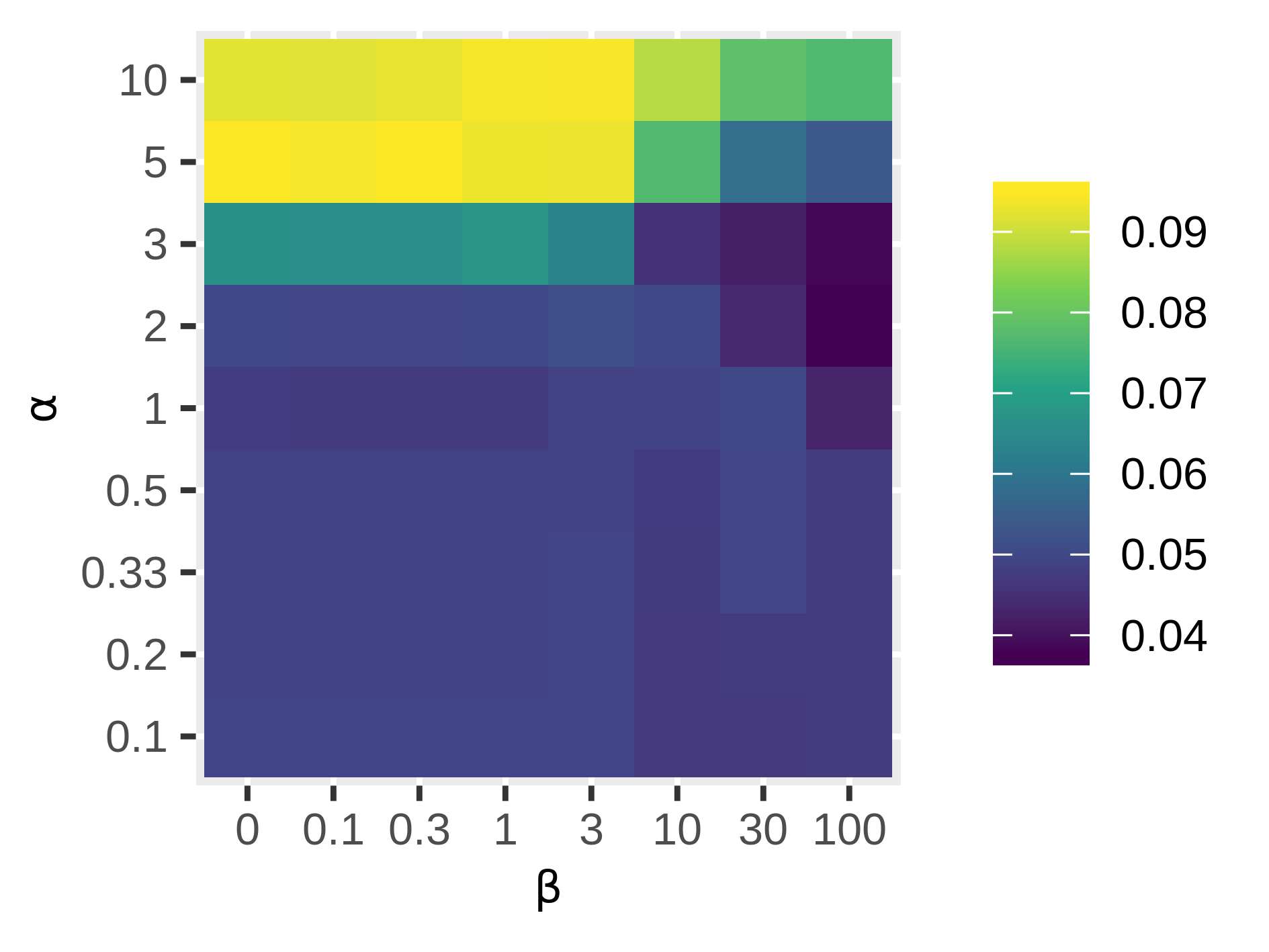}}
		\subfigure[CUBN-T (CP@100).]{\includegraphics[width=0.35\textwidth]{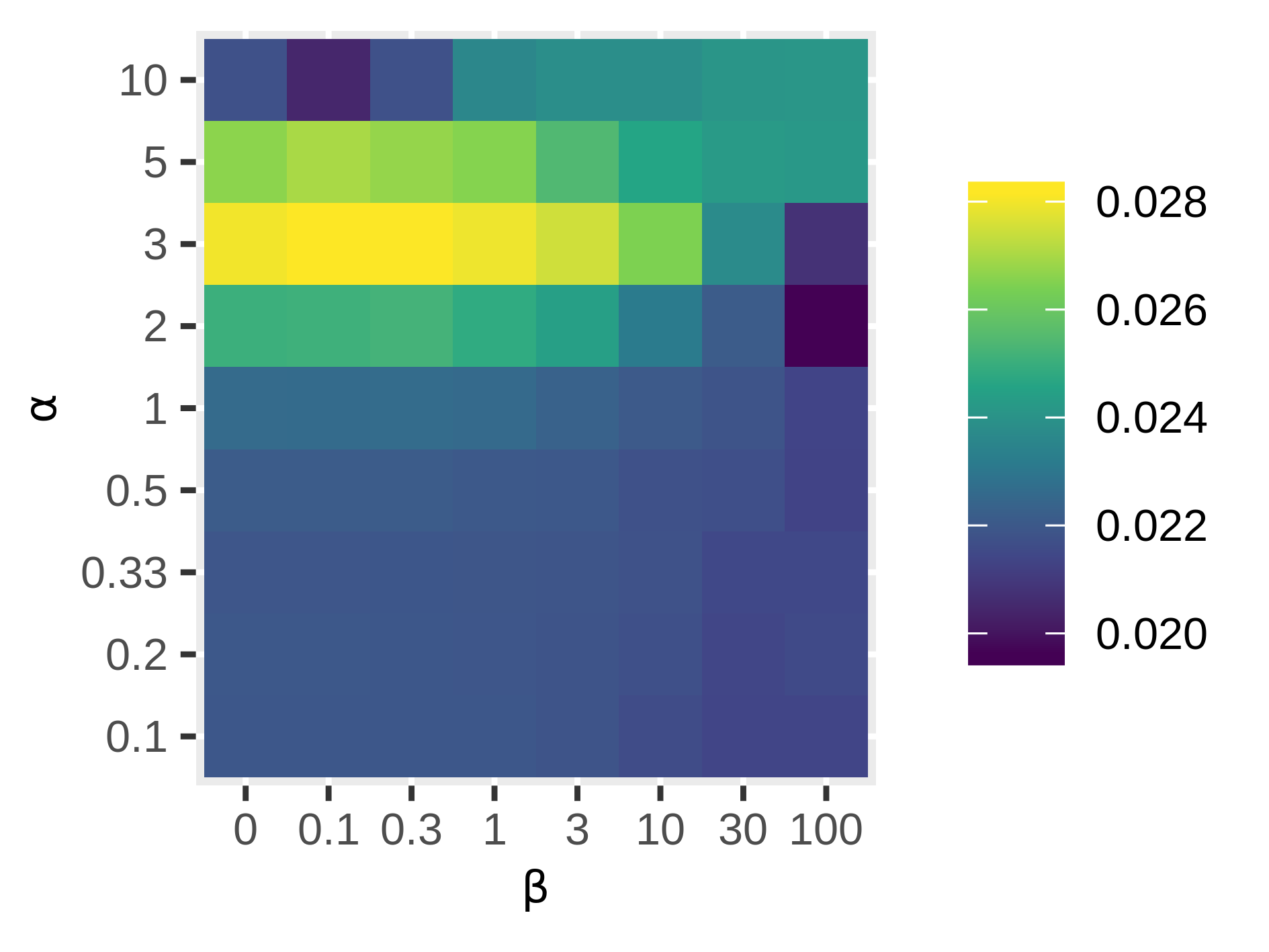}}
		\caption{Dependence on the scaling factor $\alpha$ and the shrinkage parameter $\beta$ in DH-Personalized.
			The number of neighbors are set to 2,309.}
		\label{fig:alpha_beta_dh_per}
	\end{center}
\end{figure*}

\end{document}